\documentclass[rmp,preprint,byrevtex,nobibnotes,superscriptaddress,floatfix]{revtex4}

\bibpunct[, ]{[}{]}{;}{a}{,}{,}

\usepackage{graphics}
\usepackage{graphicx}
\usepackage{psfrag}
\usepackage{latexsym}
\usepackage{textcomp}
\usepackage{amssymb}
\usepackage{amsmath}
\usepackage{bm}
\usepackage[Euler]{upgreek}
\usepackage{longtable}
\usepackage{subfigure}
\usepackage{hyperref}

\newcommand{\bfr}{\begin{flushright}}
\newcommand{\efr}{\end{flushright}}

\newcommand{\bc}{\begin{center}}
\newcommand{\ec}{\end{center}}

\newcommand{\bi}{\begin{itemize}}
\newcommand{\ei}{\end{itemize}}
\newcommand{\be}{\begin{enumerate}}
\newcommand{\ee}{\end{enumerate}}
\newcommand{\beq}{\begin{equation}}
\newcommand{\eeq}{\end{equation}}
\newcommand{\beqa}{\begin{eqnarray}}
\newcommand{\eeqa}{\end{eqnarray}}

\def\tr{\mathrm{tr} \,}

\def\vec#1{{\bf#1}}

\def\grvec#1{{\mbox{\boldmath $ #1 $} }}

\def\dyd#1{{\bf#1}}

\def\ucd#1{\vbox{\ialign{##\crcr
        \hfil$\scriptscriptstyle\mathchar"0235$\hfil\crcr
        \noalign{\kern\p@\nointerlineskip}
        $\hfil\displaystyle{#1}\hfil$\crcr}}}
\def\lcd#1{\vbox{\ialign{##\crcr
        \hfil$\scriptscriptstyle\mathchar"0235$\hfil\crcr
        \noalign{\kern\p@\nointerlineskip}
        $\hfil\displaystyle{#1}\hfil$\crcr}}}

\newcommand{\bQ}{\vec{Q}}
\newcommand{\bW}{\vec{W}}
\newcommand{\bv}{\vec{v}}
\newcommand{\bn}{\vec{n}}

\newcommand{\bM}{\dyd{M}}
\newcommand{\bk}{\bm{\kappa}}

\newcommand{\sig}{\bm{\sigma}}
\newcommand{\btau}{\bm{\tau}}

\newcommand{\avel}{{\bigl\langle}}
\newcommand{\aver}{{\bigr\rangle}}

\newcommand{\Wi}{W\!i}
\newcommand*{\average}[1]{\avel #1 \aver}

\begin{document}

\title{Coil-Stretch Transition and the Break Down  \\ of Continuum Models}
\date{\today}
\author{Mohit Bajaj}
\affiliation{Department of Chemical Engineering, Monash University,
Melbourne, Australia}
\author{Matteo Pasquali}
\affiliation{Department of Chemical and Biomolecular Engineering,
Rice University, Houston, TX~77005, USA}
\author{J. Ravi Prakash}
\email[Corresponding author: ]{ravi.jagadeeshan@eng.monash.edu.au}
\homepage[Visit: ]{http://users.monash.edu.au/~rprakash/}
\affiliation{Department of Chemical Engineering, Monash University,
Melbourne, Australia}

\begin{abstract}
The breakdown of finite element (FEM) computations for the flow  of an Oldroyd-B fluid
around a cylinder confined between parallel plates, at Weissenberg
numbers $\Wi = \mathcal{O} (1)$, is shown to
arise due to a coil-stretch transition experienced by polymer
molecules traveling along the centerline in the wake of
the cylinder. With increasing $\Wi$, the coil-stretch transition
leads to an unbounded growth in the stress maximum in the cylinder
wake. Finite element computations for a FENE-P fluid reveal that,
although polymer molecules undergo a coil-stretch transition in the
cylinder wake, the mean extension of the molecules saturates to a
value close to the fully extended length, leading to bounded
stresses with increasing $\Wi$. The existence of a coil-stretch
transition has been deduced by examining the behavior of
ultra-dilute Oldroyd-B and FENE-P fluids. In this case, the solution along 
the centerline in the cylinder wake can be obtained exactly since the velocity field
is uncoupled from the stress and conformation tensor fields. Estimation of the
number of finite elements required to achieve convergence reveals the
in-feasibility of obtaining solutions for the Oldroyd-B model for
$\Wi > 1$.

\end{abstract}

\maketitle

\section{Introduction} \label{intro}
\vspace{-0.1in}

The behavior of polymeric liquids in complex flows is intimately
linked to the distribution of molecular conformations in the flow
field. Macroscopic field variables such as the stress and velocity
are strongly coupled to microscopic quantities such as the stretch and
orientation of polymer molecules, and they influence and determine
the magnitude of each other. Recent advances in computational
rheology have led to the development of \textit{micro-macro} methods
that are capable of resolving information at various length and time
scales. However, because of computational cost, most numerical
simulations are still based on the purely macroscopic
approach of continuum mechanics, where the conservation laws of mass
and momentum are solved with a constitutive equation that relates
the stress to the deformation history, without explicitly accounting for the
microstructure~\cite{Keunings2000b}. The simplest constitutive equations
capable of capturing some qualitative aspects of the viscoelastic behavior of
polymer solutions and melts are the Oldroyd-B and upper convected
Maxwell models, respectively. These models have been
widely used in the investigation of complex flows since the
early days of computational
rheology~\cite{Owensbook}.

In spite of the apparent simplicity of the macroscopic equations,
obtaining solutions at industrially relevant values of the
Weissenberg number $W\!i$ has proven to be extremely difficult in a
range of flow geometries. Careful numerical studies over the past
few decades suggest that the principal source of
computational difficulties is the emergence of large stresses and
stress gradients within narrow regions of the flow domain.
Significant efforts have been made to develop
grid-based numerical techniques for resolving these
stresses and their gradients. In spite of considerable progress, 
numerical solutions still breakdown at
disappointingly low values of $W\!i \sim \mathcal{O} (1)$, and it is
still not clear whether this is because solutions do not exist at
higher values of $W\!i$, or whether it is simply due to the
inadequacy of current numerical techniques~\cite{Keunings2000b}.
Very recently, \citet{Renardy2006}  has shown analytically
that in the special case of steady flows with an \textit{interior
}stagnation point, the mathematical structure of the upper convected
Maxwell and Oldroyd-B  models can be expected to lead to
singularities in the viscoelastic stresses and their gradients with
increasing $W\!i$.

Nearly two decades ago, \citet{Hinch1988} argued in a seminal paper
that in the case of the Oldroyd-B model,  the inability to
compute macroscopic flows  at high Weissenberg
numbers (the so called high Weissenberg number problem, or HWNP),
has a physical origin in a microscopic phenomenon. 
The Oldroyd-B model predicts an unbounded extensional viscosity in
homogenous extensional flows at a critical value of $W\!i$. 
The Oldroyd-B constitutive equation can be derived from kinetic
theory by representing polymer molecules by 
Hookean dumbbells. The unphysical behavior in extensional flows is
due to the infinite extensibility of the Hookean spring used in the
model. By considering the simple example of a stagnation point flow
of an Oldroyd-B fluid, Rallison and Hinch showed that when the
strain rate is supercritical, infinite stresses can occur in the
interior of a steady flow, brought about by the unbounded
stretching of polymer molecules. Based on their analysis, they
suggested the use of a constitutive equation that is derived from a
microscopic model with a nonlinear spring force law (which would
impose a finite limit on a polymers extension), as an obvious remedy
for the HWNP.

\citet{Chilcott1988}  examined the benchmark complex flow problems
of unbounded flow around a cylinder and a sphere, using a dumbbell
model with finite extensibility, as a means of demonstrating the
validity of this analysis. In order to understand the coupling
between the polymer extension by flow, the stresses developed in the
fluid,  and the resultant flow field, they deliberately used the
conformation tensor as the fundamental variable instead of the
stress. The conformation tensor gives information on the
distribution of polymer conformations within the flow field in an
averaged sense. The use by Chilcott and Rallison of kinetic theory
to develop their model  enabled the derivation of a simple
expression relating the conformation tensor to the polymer
contribution to the stress. By solving the equation for  the
conformation tensor along with the mass and momentum conservation
laws, Chilcott and Rallison showed that even though there existed
highly extended material close to the boundary and in the wake of the obstacle, 
there no longer was an upper limit to
$W\!i$ in the range of values that was examined in their
computations. Since the degree of molecular extension is directly
related to the magnitude of stress, the Chilcott and Rallison
procedure established a clear connection between high stresses and
stress gradients in the flow domain with the configurational and
spatial distribution of polymer conformations. Indeed, when
simulations were carried out with the polymer length set to infinity
rather than a finite value, the downstream structure was no longer
resolvable, and the mean stretch of the polymers in the flow
direction continued to grow with increasing $W\!i$ until the
solution failed.

\begin{figure}
\psfrag{Z}{\large  $\bv_{\bf x}={\bf f(y)}$} \psfrag{A}{\large
$\bv_{\bf y} ={\bf 0}$} \psfrag{B}{\large $\vec{v}\cdot
\grvec{\nabla}{\dyd{M}}=\vec{0}$} \psfrag{F}{\large $\bv_{\bf
x}={\bf f(y)}$} \psfrag{G}{\large $\bv_{\bf y} ={\bf 0}$}
\psfrag{M}{\large $\vec{v}=\vec{0}$} \psfrag{D}{\large
$\vec{t}\vec{n}:{( \btau_\text{s}+ \sig)}=0$} \psfrag{E}{\large
$\bv_{\bf y} ={\bf 0}$} \resizebox{16.0cm}{!}
{\includegraphics*{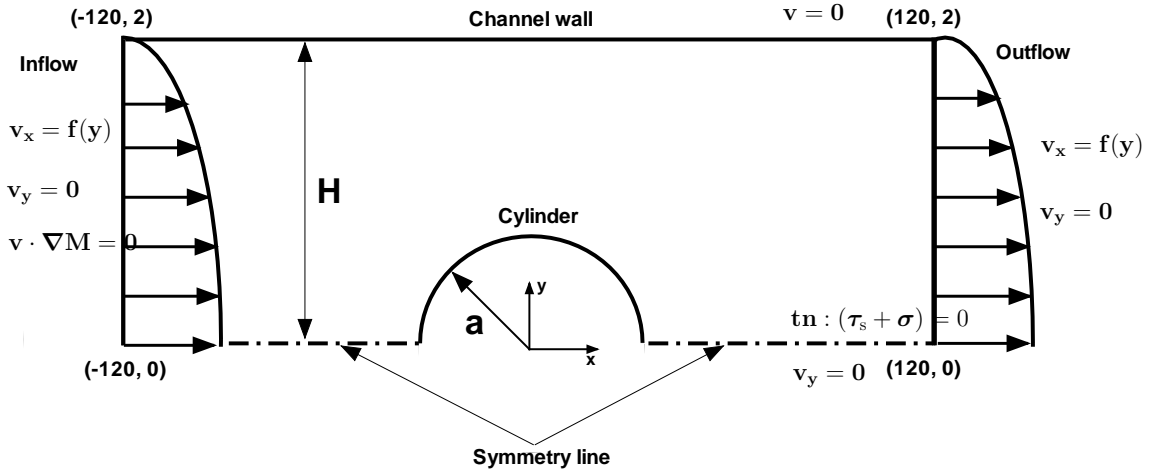}} \caption{Flow domain and
boundary conditions for the flow around a cylinder confined between
parallel plates.} \label{Flow-Geometry}
\end{figure}

In spite of this compelling demonstration of the physical origin of
the HWNP, the upper convected Maxwell and Oldroyd-B  models have
continued to be used extensively in computational rheology. The
reason for this might perhaps be attributed to the fact that even
though stresses may be large, they are still bounded, and so far,
there has been no conclusive demonstration that bounded solutions
for the viscoelastic stress do not exist in complex flows at high
values of $W\!i$. On the contrary, by considering the steady flow of
upper convected Maxwell and Oldroyd-B fluids around a cylinder
confined between two parallel plates (a  schematic of the flow
geometry is displayed in Fig.~\ref{Flow-Geometry}),
\citet{Renardy2005}  have presented strong numerical evidence that
suggests that solutions do exist for $W\!i > 1$, and that current
numerical techniques are not able to resolve them.

The particular  benchmark problem of flow around a cylinder between
parallel plates was chosen by these authors since even though the
geometry has no singularities, the maximum values of  $W\!i$ for
which converged solutions exist are amongst the smallest of all
benchmark
flows~\cite{Fan1999,Sun1999,Alves2001,Caola2001,Oscar2006}. The
presence of upstream and downstream stagnation points leads to the
development of steep stress boundary layers near the cylinder and in
the wake of the cylinder, making the flow a stringent test of any
numerical technique. Rather than solving the coupled problem
simultaneously for the velocity and conformation tensor fields,  for
which the existence of solutions at high  $W\!i$ is unknown,
Wapperom and Renardy assumed a \textit{Newtonian-like} velocity
field and solved only for the conformation tensor field. The chosen
velocity field has an analytical representation, is very similar to
the Newtonian velocity field in the same geometry (and the velocity
field for an Oldroyd-B fluid at $W\!i \sim \mathcal{O} (1)$), and
satisfies all the key requirements for the velocity field near the
cylinder. The advantage of this approach is that the existence of a
solution for an upper convected Maxwell model in such a velocity
field, at all values of $W\!i$ is guaranteed~\cite{Renardy2000}, and
hence failure of a numerical scheme can be attributed purely to
numerics. By developing a Lagrangian technique which involves
integrating the conformation tensor equation along streamlines using
a predictor-corrector method, the authors were able to compute
stresses up to arbitrarily large values of $W\!i$ (as high as 1024),
and as a result, establish conclusively the existence of narrow
regions with very high stresses near the cylinder, and in its wake.
Further, they show that although the velocity field is known,  one
of the currently used numerical techniques, the backward-tracking
Lagrangian technique, is unable to resolve the extremely thin stress
boundary layers even for relatively low values of $W\!i$. Since the
fully coupled problem and the fixed flow kinematics problem share
the same basic dilemma of computing the stress field, Wapperom and
Renardy argue that solutions also probably exist for $W\!i > 1$ for
the steady flow of Oldroyd-B and upper convected Maxwell fluids
around a cylinder confined between parallel plates, but current
numerical techniques are not able to resolve them.

The use of the conformation tensor as the fundamental quantity
rather than the stress has become common in computational rheology,
and the challenge of developing numerical methods capable of
resolving steep stresses and stress gradients has been transformed
to one of developing techniques capable of resolving rapidly varying
conformation tensor fields. In an important recent breakthrough,
\citet{Fattal2004} have shown that by changing the fundamental
variable to the matrix logarithm of the conformation tensor, stable
numerical solutions can be obtained at values of $W\!i$
significantly greater than ever obtained before. The success of
their variable transformation protocol is predicated on their
identification of the source of the HWNP as the inability of methods
based on polynomial basis functions (such as finite element
methods), to adequately represent the exponential profiles that
emerge in conformational tensor fields in the vicinity of stagnation
points and in regions of high deformation rate.

\citet{Fattal2005a} have recently carried out a stringent test of
the log conformation representation by examining the flow of
Oldroyd-B and Giesekus fluids around a cylinder confined between
parallel plates. (It is worth noting that unlike the Oldroyd-B
models prediction of unbounded extensional viscosity at a finite
extension rate, the extensional viscosity predicted by the Giesekus
model is always finite~\cite{BAH1}). For both
the fluids, \citet{Fattal2005a} find that with the log
conformation formulation, the solution remains numerically stable
for values of $W\!i$ considerably greater than those obtained
previously with standard finite element (FEM) implementations. On
the other hand, the two fluids differ significantly from each other
with regard to the behavior of the convergence of solutions with
mesh refinement. The lack of convergence with mesh refinement is
usually seen most dramatically in the failure of different meshes to
accurately predict the maximum that occurs in  the normal polymeric
stress component ${\sigma}_{xx}$  on the centerline in the wake of
the cylinder. In the case of the Giesekus model, even at values of
Weissenberg number as high as $W\!i = 100$, mesh convergence is
achieved in large parts of the flow domain,  with the exception of
localized regions near the stress maximum in the wake where
convergence is not realized. For the Oldroyd-B model, however, the
log conformation formulation fails to achieve mesh convergence in
the entire wake region at roughly the same value ($W\!i \gtrsim
0.6$) as in previous studies. Further, the solution becomes unsteady
at some greater value of $W\!i$ (depending on the mesh), finally
breaking down at even higher  $W\!i$. \citet{Fattal2005a}  
speculate that this failure is probably due to the infinite
extensibility of the Hookean dumbbell model that underlies the
Oldroyd-B model, and call for further investigations to see if this
might lead to the non-existence of solutions beyond some value of
$W\!i$. Thus, after many years of attempting to resolve the HWNP by
purely numerical means, it's origin still remains a mystery.

In this paper, we conclusively establish the connection between the
HWNP and the unphysical behavior of the Oldroyd-B model, in the
benchmark problem of the steady flow around a cylinder confined
between parallel plates. We show that when $W\!i \approx 1$, polymer
molecules flowing along the centerline in the wake of the cylinder
undergo a \emph{coil-stretch} transition, and that the location of
the transition coincides with the maximum in the normal stress
component ${\sigma}_{xx}$ on the centerline. With increasing $W\!i$,
the molecules stretch without bound, and this is accompanied by a
normal stress that increases without bound. (For an UCM fluid,
\citet{Alves2001}   have previously speculated that ${\sigma}_{xx}$
may develop a singularity at the position where it reaches a maximum
value, based on the asymptotic behavior of highly accurate numerical
results obtained by them using a finite volume method. However, they
were unable to attribute a precise reason for the occurrence of a
singularity). Computations carried out here for a FENE-P fluid
reveal that in this case also, polymer molecules undergo a
coil-stretch transition which is located at the stress maximum.
However, with increasing  $W\!i$, the mean extension of the
molecules saturates to a value close to their fully extended length,
enabling computations beyond the critical Weissenberg number.

These \textit{dilute solution} results have been obtained by drawing
on insights gained from the solution of an \textit{ultra-dilute}
model, in which the velocity field is decoupled from the
conformation tensor (and stress) field. This procedure is similar to
the earlier work by \citet{Renardy2005}  described above. However,
rather than using an ad hoc velocity field, we have solved for the
Newtonian velocity field using a full-fledged FEM simulation. Even
for the ultra-dilute model, with a known velocity field, the
solution of the equation for the conformation tensor  (which is
denoted here by $\bM$) for the Oldroyd-B model is found to breakdown
for  $W\!i \sim \mathcal{O} (1)$, when a standard FEM method is
used. However, an exact (numerical) solution for $\bM$, valid for
arbitrary large values of $W\!i$, is obtained along the centerline
using \emph{two} different techniques.  In the first method, we
exploit the fact that the equation for $\bM$ in the Oldroyd-B and
FENE-P  models reduces at steady state to a system of ordinary
differential equations (ODEs) along the centerline. In the second
method, trajectories of the end-to-end vectors of an ensemble of
dumbbells, flowing down the centerline in the wake of the cylinder,
are calculated by carrying out Brownian dynamics simulations using
the known velocity field. Averages carried out over the ensemble of
trajectories lead to macroscopic predictions that are identical to
the  exact (numerical) results obtained by solving the macroscopic
model ODEs discussed above, for arbitrary values of $W\!i$.
Comparison of the FEM results with the exact numerical results
enables a careful examination of the reasons for the breakdown of
the finite element method. In particular, the occurrence of a
coil-stretch transition at the location of the stress maximum in the
wake is clearly demonstrated, and an estimation of the number of
elements in the FEM method required to achieve convergence with
increasing  $W\!i$ reveals the in-feasibility of obtaining solutions
for $W\!i > 1$.

The plan of this paper is as follows. In section~\ref{dilsol}, we
summarize the governing equations, boundary conditions and
computational method for the flow of dilute Oldroyd-B and FENE-P
fluids around a cylinder confined between parallel plates. We also
elaborate on the connection between the macroscopic models, and the
Kinetic theory models from which they are derived. In
section~\ref{ultdil}, we describe the means by which exact solutions
to the ultra-dilute models may be obtained, and examine the nature
of the maximum in the $\text{M}_{xx}$ component of the conformation
tensor. The results of FEM computations and the exact numerical
methods are first discussed for ultra-dilute solutions in
section~\ref{RandD}, followed by a discussion of the predictions of
FEM computations for the dilute model. Section~\ref{conc} summarizes
the main conclusions of this work.

\section{Dilute Solutions} \label{dilsol}

\subsection{Basic equations} \label{baseqn}

As displayed in Fig.~\ref{Flow-Geometry}, the cylinder axis is in
the $z$-direction perpendicular to the plane of flow. With the
assumption of a plane of symmetry along the centreline ($y=0$),
computations are only carried out in half the domain. The cylinder,
with radius $a$, is assumed to be placed exactly midway between the
plates, which are separated from each other by a distance $2H$. In
common with other benchmark flow around a confined cylinder
simulations, we set the \emph{blockage} ratio $H/a = 2$.

We normalize all macroscopic length scales with respect to $a$,
velocities with respect to the mean inflow velocity far upstream
$\average v $, macroscopic time scales with respect to $ a /
\average v$, and stresses and pressure with respect to $\eta \,
{\average v} /a$, where $\eta = \eta_\text{s} + \eta_{\text{p,0}}$
is the sum of the Newtonian solvent viscosity $\eta_\text{s}$ and
the zero-shear rate polymer contribution to viscosity
$\eta_{\text{p,0}}$. Microscopic length and time scales are
discussed subsequently. The two dimensionless numbers of relevance
here are the Weissenberg number $W\!i = \lambda \, {\average v} /a$,
in which $\lambda$ is a relaxation time, and the Reynolds number $Re
= \rho \, a \, {\average v} /\eta$, where $\rho$ is the fluid
density.

The complete set of non-dimensional governing equations for a dilute
polymer solution, described by the Oldroyd-B or FENE-P models, is
\begin{flalign}
\vec{\nabla} \cdot \vec{v} &=0 && \text{\small \it (Mass balance)} \label{mass}\\
Re \, \vec{v} \cdot \vec{\nabla}\vec{v}- \vec{\nabla} p -
\vec{\nabla} \cdot \btau_\text{s}
- \vec{\nabla} \cdot \sig  &=\vec{0}  && \text{\small \it (Momentum balance)}\label{mom}\\
\frac{\partial \dyd{M}}{\partial t} + \vec{v} \cdot \nabla\dyd{M} -
\bk \cdot \dyd{M} -  \dyd{M} \cdot \bk^T & = -
\frac{1}{W\!i}\left\lbrace f (\tr \dyd{M}) \, \dyd{M}   - \dyd{I}
\right\rbrace &&\text{\small \it
(Conformation tensor)}\label{conf}\\
\btau_\text{s} & = 2 \, \beta \, \dyd{D} &&\text{\small \it
(Solvent stress)}\label{solstr}\\
\sig & = \frac{(1 - \beta)}{W\!i}\left\lbrace f (\tr \dyd{M}) \,
\dyd{M} - \dyd{I} \right\rbrace   &&\text{\small \it (Polymer
stress)}\label{polstr}
\end{flalign}
\noindent In these equations, $\bk = (\nabla \vec{v})^T$ is the
transpose of the velocity gradient, $\dyd{D}  =
\frac{1}{2}\left(\bk^T +\bk \right)$ is the rate of deformation
tensor, and the parameter $\beta = (\eta_\text{s} / \eta) $ is  the
viscosity ratio. Here, we use $\beta=0.59$, and set $Re=0$, which
are the values used in benchmarks for the Oldroyd-B model. The form
of the function $ f (\tr \dyd{M})$ depends on the microscopic model
used to derive the equation for the evolution of the conformation
tensor, and is consequently different in the Oldroyd-B and FENE-P
models, as elaborated below.

\subsubsection{Oldroyd-B model} \label{oldB}

As mentioned earlier, the Oldroyd-B constitutive equation can be
derived from Kinetic theory, with the polymer molecule represented
by a Hookean dumbbell model consisting of two beads connected
together by a spring~\cite{BAH2}. The spring obeys a linear spring
force law $\vec{F}^\text{(s)} = H \bQ$, where $H$ is the spring
constant, and $\bQ$ is the connector vector between the beads. In
this model, the conformation tensor $\dyd{M}$ is defined by the
expression,
\begin{equation}
\dyd{M} = \frac{1}{(\average{Q^2}_\text{eq}/3)}\, \average{\bQ \bQ}
\label{conf2}
\end{equation}
where, $\average{( .)}$ denotes an ensemble average, and
$\sqrt{\average{Q^2}_\text{eq}}$ is the root mean square end-to-end
vector at equilibrium. Note that, $\sqrt{\average{Q^2}_\text{eq}/3}
= \sqrt{ k_\text{B} T / H}$ (with $k_\text{B}$ being the Boltzmanns
constant and  $T$ the temperature), is the microscopic length scale.
The connection between the microscopic and macroscopic models is
established by deriving an evolution equation for the conformation
tensor, and by relating the macroscopic polymeric stress to the
conformation tensor. The expression of the conformation tensor is
given by eqn~(\ref{conf}) above, where in this case, the function
$f$ is identically equal to unity, {\it i.e.}\ $ f (\tr \dyd{M}) =
1$. The polymer contribution to the stress is given by Kramers
expression~\cite{BAH2},
\begin{equation}
\sig  = n \, k_\text{B} T \, \left\lbrace  \dyd{M} - \dyd{I}
\right\rbrace \label{polstrob}
\end{equation}
 where, $n$ is the number density of polymer molecules in solution.
With the relaxation time $\lambda$, and the zero-shear rate polymer
contribution to the viscosity $\eta_\text{p,0}$, of the Oldroyd-B
model related to microscopic model parameters by,
\begin{equation}
\lambda  = \frac{\zeta}{4H} \,; \quad \eta_\text{p,0}  = n
k_\text{B} T \, \lambda
\end{equation}
where, $\zeta$ is the bead friction coefficient, it is
straightforward to see that the non-dimensionalization scheme used
here leads to  equation~(\ref{polstr}) for $\sig$.

\subsubsection{FENE-P model} \label{fenep}

The FENE-P model corrects the failing of the Hookean dumbbell model
with its infinite extensibility, by using a spring force law that
ensures that the magnitude of the end-to-end vector remains (on an
average) below the fully extensible length of the spring $Q_0$,
\begin{equation}
\vec{F}^\text{(s)} = \frac{H}{\left( 1 -{\average{Q^2}} / {Q_0^2}
\right)}  \, \bQ
\end{equation}
For the FENE-P model, the conformation tensor is also defined by
eqn~(\ref{conf2}), but with the mean square end-to-end vector at
equilibrium ${\average{Q^2}_\text{eq}} $ given by,
\begin{equation}
\average{Q^2}_\text{eq} = \frac{3 \, Q_0^2 \, ( k_\text{B} T /
H)}{Q_0^2 + 3 \, (k_\text{B} T / H)}
\end{equation}
Defining the \emph{finite extensibility} parameter $b$ by,
\begin{equation} \label{finext}
b = \frac{ Q_0^2 }{\average{Q^2}_\text{eq}}
\end{equation}
the function $ f (\tr \dyd{M})$ for the FENE-P model in the
conformation tensor evolution equation~(\ref{conf}) can be shown to
be given by~\cite{Pasquali2002a},
 \begin{equation} \label{funcf}
f(\tr \dyd{M}) = \frac{b - 1}{b - \tr \dyd{M} /3}
\end{equation}
Note that the definition of $b$ used in~eqn~(\ref{finext}) is
different from the definition of the finite extensibility parameter
given in~\citet{BAH2}, which is widely used in the literature. In
the Bird \emph{et al.}\  definition, the microscopic length scale
used to non-dimensionalize $Q_0$ is the Hookean dumbbell length
scale $\sqrt{k_\text{B} T / H}$. The polymeric stress for a FENE-P
fluid is given by,
\begin{equation}
\sig  = n \, k_\text{B} T \, \left\lbrace f (\tr \dyd{M}) \, \dyd{M}
- \dyd{I} \right\rbrace \label{polstrfene}
\end{equation}
With the relaxation time $\lambda$, and the zero-shear rate polymer
contribution to the viscosity $\eta_\text{p,0}$ related to
microscopic model parameters by,
\begin{equation}
\lambda  =\left(\frac{b - 1}{b } \right)  \frac{\zeta}{4H} \,; \quad
\eta_\text{p,0}  = n k_\text{B} T \, \lambda
\end{equation}
non-dimensionalization of eqn~(\ref{polstrfene})  leads to
eqn~(\ref{polstr}).

\begin{table}[t]
\begin{center}
\caption{Meshes used in the finite element simulations for computing
viscoelastic flow around a cylinder.} \label{tab:Mesh}
    \vspace{0.1in}
 \begin{tabular}{|c|c|c|c|c|c|} \hline
 Mesh & M1  & M2  & M3 & M4 & M5 \\\hline
 Number of Elements & 2311 & 5040 & 8512 & 16425 &
31500 \\ \hline
\end{tabular}
\end{center}
\end{table}

\begin{figure}[tbp]
\begin{center}
\resizebox{14.0cm}{!}{\includegraphics{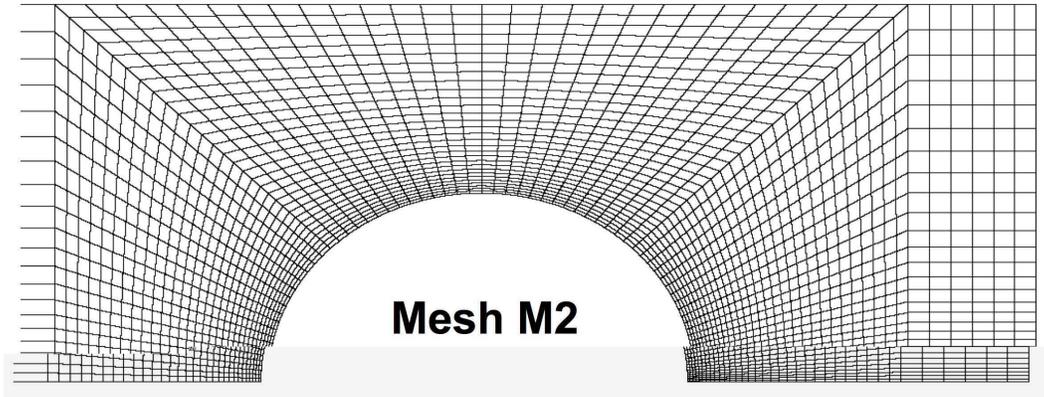}}
\end{center}
\caption{Mesh M2 used in the finite element simulations.}
\label{meshM2}
\end{figure}

\subsection{Boundary conditions and computational method} \label{bound}

The set of governing equations~(\ref{mass})-(\ref{polstr})  are
solved with the boundary conditions shown in
Fig.~\ref{Flow-Geometry}. The location of the inflow and outflow
boundaries coincides with that chosen by~\citet{Sun1999}, who showed
that the flow is insensitive to further displacement of the open
boundaries in the range of Weissenberg numbers examined. A no-slip
boundary condition is imposed on the cylinder surface and the
channel walls. Fully developed flow is assumed at both inflow and
outflow boundaries, with the velocity prescribed at both boundaries.
Here, $\average v = 1$ for the prescribed velocity field. The
boundary conditions on the conformation tensor are imposed only at
the inflow boundary. At the symmetry line, $\vec{t}\vec{n}:{(
\btau_\text{s}+ \sig)}=0$ and $v_y = 0$ is imposed, where, $\vec{t}$
and $\bn$ are the unit vectors tangential and normal to the symmetry
line, respectively.

The governing equations are discretized by using the DEVSS-TG/SUPG
mixed finite element method~\cite{Pasquali2002a}.  The DEVSS-TG
formulation involves the introduction of an additional variable, the
velocity gradient $\bk$. Continuous biquadratic basis functions are
used to represent velocity, linear discontinuous basis functions to
represent pressure and continuous bilinear basis functions are used
for the interpolated velocity gradient and conformation tensor. The
DEVSS-TG/SUPG spatial discretization results in a large set of
coupled non-linear algebraic equations, which are solved by Newton's
method with analytical Jacobian and first order arc-length
continuation in $W\!i$~\cite{Pasquali2002a}. Five different meshes
are used for the FEM calculations. Details of the different meshes
used in this work are given in Table~\ref{tab:Mesh}, and the mesh M2
is displayed in Fig.~\ref{meshM2}.  The important distinction among
the five meshes is the density of elements on the cylinder surface
and in the wake of the cylinder.

\section{Ultra-Dilute Solutions} \label{ultdil}

The phrase ``ultra-dilute" is used to describe a situation where,
even though polymer molecules are present, they have a negligible
effect on the velocity field. As demonstrated in the earlier work by
\citet{Renardy2005} , the solution of the ultra-dilute problem
provides important insights into the structure of the solution to
the dilute problem, where the polymer stress and velocity fields are
fully coupled. As will be clear from the results and discussion
presented subsequently, the solution of the ultra-dilute case lies
at the heart of the analysis carried out in this work.

Since the velocity field for an ultra-dilute solution is determined
completely by the solvent stress, it is identical to the velocity
field for a Newtonian fluid. The ultra-dilute conformation tensor
and velocity fields are simply obtained by solving the governing
equations with the parameter $\beta$ set equal to unity. As can be
seen from eqn~(\ref{polstr}), this implies $\sig=0$, leading to
eqns~(\ref{mass}) and (\ref{mom}) being identical to the mass and
momentum balances for a Newtonian fluid. The same FEM formulation
described above for the solution of the dilute case, can
consequently,  also be used to obtain the ultra-dilute conformation
tensor and Newtonian velocity fields when $\beta=1$.

The conformation tensor field, for an ultra-dilute solution,
corresponds to the average configurations of polymer molecules in a
pre-determined velocity field. Note that the choice of value for
$W\!i$ has no influence on the velocity field (which is determined
once and for all for the specified geometry), but can significantly
effect the conformation tensor field, as will be seen subsequently.

Clearly, the polymer contribution to the stress tensor cannot be
obtained from a solution of the macroscopic equations, where it is
assumed to be zero. In this case, we resort to reporting the
dimensionless stress predicted by the microscopic models, $\sig^* =
\sig/ n \, k_\text{B} T = (\lambda/\eta_{\text{p,0}})\, \sig$, given
by eqns~(\ref{polstrob}) and (\ref{polstrfene}), in the Oldroyd-B
and FENE-P models, respectively. This is similar to the evaluation
of the polymer contribution to the stress in a homogenous simple
shear or extensional flow, where the velocity field is prescribed a
priori.

As mentioned earlier, the FEM formulation used here fails to give
mesh-converged results for the conformation tensor field for an
ultra-dilute solution beyond a threshold value of the Weissenberg
number. The reasons for this failure are analyzed subsequently.
Crucial for this analysis, however, is the possibility of obtaining
an exact (numerical) solution to the conformation tensor equation
along the centerline in the wake of the cylinder. Two means of
obtaining  such a solution are discussed below.

\subsection{Exact (numerical) solution along the centerline in the cylinder wake} \label{exacsol}

\subsubsection{System of ODEs} \label{ode}

By the requirements of symmetry along the centerline, the velocity
field must have the form, $v_x = g(x)$, $v_y = 0$,  and the
components of the velocity gradient tensor must satisfy,
$\kappa_{xy} = \kappa_{yx} = 0$. Incompressibility requires
$\kappa_{xx} = - \kappa_{yy}$. The flow along the centerline is
consequently planar extensional in character.

Substituting these results into the evolution equation for the
conformation tensor~(eqn~(\ref{conf})) leads, at steady state, to a
system of ODEs in the independent variable $x$, for the components
of the conformation tensor $\dyd{M}$. In the case of the Oldroyd-B
model, the equations for each of the components are decoupled from
each other. For the purposes of analysis in the present work,  we
are only interested in the equation for the $\text{M}_{xx}$
component, which can be shown to be,
\begin{equation}\label{old-B}
\frac{d\text{M}_{xx}}{dx}
=-\frac{(1-2\lambda\kappa_{xx}(x))}{\lambda v_{x}(x)} \,
\text{M}_{xx}+\frac{1}{\lambda v_{x}(x)}
\end{equation}

Since the velocity field is known a priori, equation~(\ref{old-B})
is a first order liner ODE for $\text{M}_{xx}$ that can be solved
straight-forwardly once a boundary condition is prescribed. We can
expect that at the downstream stagnation point at $x=1$, the Hookean
dumbbells that represent the polymer molecules are in their
equilibrium configurations, and consequently $\text{M}_{xx}=1$.
However, as can be seen from the form of eqn~(\ref{old-B}), we
cannot use this boundary condition since $v_x=0$ at the stagnation
point. We can overcome this difficulty by exploiting the fact that
we can guess the form of the velocity field  asymptotically close to
the stagnation point. For an unbounded flow of a Newtonian fluid
near a stagnation point on the surface of a two-dimensional body,
the assumption of a quadratic velocity field, asymptotically close
to the stagnation point, is exactly valid for Stokes flow, and is
consistent with the accepted numerical solution for Hiemenz
flow~\cite{pozrikidisbook}. The approximate Newtonian velocity field
postulated by~\citet{Renardy2005}  for the flow around a confined
cylinder is also of the form  $v_x= k \, (x-1)^2$ (see
also~\cite{Renardy2000}), with $k=4$, in the limit as $x \to 1$
(from above). We find that the assumption of a quadratic velocity
field, with a value of $k=4.178$, leads to an excellent fit of the
Newtonian velocity field close to the downstream stagnation point
obtained by the FEM solution. For a quadratic velocity field,
eqn~(\ref{old-B}) admits an analytical solution for $\text{M}_{xx}$,
\begin{equation}\label{mxxanal}
\text{M}_{xx}=1+2\alpha \, (x-1)+3\alpha^2 \, (x-1)^2+3\alpha^3 \,
(x-1)^3+1.5\alpha^4 \, (x-1)^4
\end{equation}
where, $\alpha = 2 k \lambda$. As a result, we use the analytical
value of $\text{M}_{xx}$ at $x=1.01$ as the boundary condition to
integrate  eqn~(\ref{old-B}), with a Runge-Kutta 4th order method.
The functions $v_x(x)$ and $\kappa_{xx} (x) $ are obtained by
interpolation from the Newtonian FEM solution, at each of the values
of $x$ where they are required for the purpose of integration.

For the FENE-P model, the diagonal components are not decoupled from
each other, and consequently, evaluation of the $\text{M}_{xx}$
component requires a solution of a system of ODEs for all the
diagonal components, as can be seen from the equations below,
\begin{equation}
\begin{aligned}
\frac{d \text{M}_{xx}}{dx} &=-\frac{1}{\lambda v_{x}(x)} \,
\left\lbrace \frac{b-1}{b-\tr \dyd{M}/3} - 2\lambda\kappa_{xx}(x)
\right\rbrace \, \text{M}_{xx} + \frac{1}{\lambda v_{x}(x)} \\
\frac{d \text{M}_{yy}}{dx} &=-\frac{1}{\lambda v_{x}(x)} \,
\left\lbrace \frac{b-1}{b-\tr \dyd{M}/3} - 2\lambda\kappa_{yy}(x)
\right\rbrace \, \text{M}_{xx} +\frac{1}{\lambda v_{x}(x)} \\
\frac{d \text{M}_{zz}}{dx} &=-\frac{1}{\lambda v_{x}(x)} \,
\left\lbrace \frac{b-1}{b-\tr \dyd{M}/3} \right\rbrace \,
\text{M}_{zz} +\frac{1}{\lambda v_{x}(x)}  \label{FENE-P}
\end{aligned}
\end{equation}
It is difficult to solve these equations analytically even with the
assumption of a quadratic velocity field close to the stagnation
point. Since $\kappa_{xx} (x) \ll 1$ at $x=1.01$,  we use the
equilibrium initial conditions, $\text{M}_{xx} =1$, $\text{M}_{yy}
=1$, and $\text{M}_{zz} =1$, as the boundary conditions, though
these values are strictly correct only at the stagnation point
$x=1$. It turns out, however, that the conformational tensor fields
along the centerline, downstream of the stagnation point, are
insensitive  to a variation  by a few percent, in the boundary
values chosen for the diagonal components of $\dyd{M}$ close to the
stagnation point. (For instance, identical results are obtained if
the boundary conditions for the Oldroyd-B model at $x=1.01$, are
used instead).

\subsubsection{Brownian dynamics simulations} \label{bds}

An alternative means of obtaining an exact solution along the
centerline in the cylinder wake is to exploit the connection between
the macroscopic and microscopic models. If we imagine an ensemble of
dumbbells at any position $x$ on the centerline, subject to the
local velocity gradient, we expect that at steady state, the
ensemble average $\average{\bQ^\dagger \bQ^\dagger} = \dyd{M}$,
where, $\bQ^\dagger = \bQ / \sqrt{(\average{Q^2}_\text{eq}/3)}$. We
could also imagine a packet of fluid with an ensemble of dumbbells,
starting close to the stagnation point and traveling down the
centerline with velocity $v_x(x)$, experiencing the local velocity
gradient at each position $x$. In this case,  the variation with
time of $\average{\bQ^\dagger \bQ^\dagger} $ would be equivalent to
the variation of $\dyd{M}$ with $x$ in the macroscopic models. The
ensemble average $\average{\bQ^\dagger \bQ^\dagger}$  can be
obtained by integrating the stochastic differential equation (SDE),
\begin{equation} \label{sde}
d \bQ^\dagger = \lbrace \bk(t) \cdot \bQ^\dagger - \frac{1}{2W\!i}
\, f \left(\average{{Q^\dagger}^2} \right) \bQ^\dagger \rbrace \, dt
+ \frac{1}{\sqrt{W\!i}} \, d \bW^\dagger
\end{equation}
which governs the stochastic dynamics of Hookean or FENE-P dumbbells
subject to the time varying velocity gradient $\bk
(t)$~\cite{Ottingerbook}. Here, $\bW^\dagger$ is a non-dimensional
Wiener process, and
 \begin{equation}
f \left(\average{{Q^\dagger}^2} \right)  = \begin{cases}
1  & \text{for Hookean dumbbells}, \\
({b - 1})/ ({b - \average{{Q^\dagger}^2} /3}) & \text{for FENE-P
dumbbells}
 \end{cases}
\end{equation}
is the same function for the FENE-P model as in eqn~(\ref{funcf}),
with $\average{{Q^\dagger}^2}$ taking the place of $\tr \dyd{M}$.

The only remaining issue is to obtain $\kappa_{xx} (t)$ for a packet
of fluid traveling down the centerline in the wake of the cylinder.
This can be done in a straightforward way since we know $v_x (x)$
and $\kappa_{xx} (x)$ for a Newtonian fluid.  Since $dx/dt =
v_x(x)$, the integral,
\begin{equation} \label{tvsx}
t = \int_{1+\delta}^x dx^\prime \, \frac{1}{v_x (x^\prime)} \equiv
h(x)
\end{equation}
gives us $t$ as a function of $x$ for a material particle. Clearly,
$\kappa_{xx}(t) = \kappa_{xx}( h^{-1} (t))$. The fluid packet would
have to start its journey slightly downstream of the stagnation
point (represented by $1+ \delta$ in the lower limit of the
integral in  eqn~(\ref{tvsx})), since otherwise it would remain
indefinitely at the stagnation point. Time is therefore initialized
at $x = 1+ \delta$.
\begin{figure}[tbp]
\begin{center}
\includegraphics[bb= 86 261 508 580, scale=0.75]{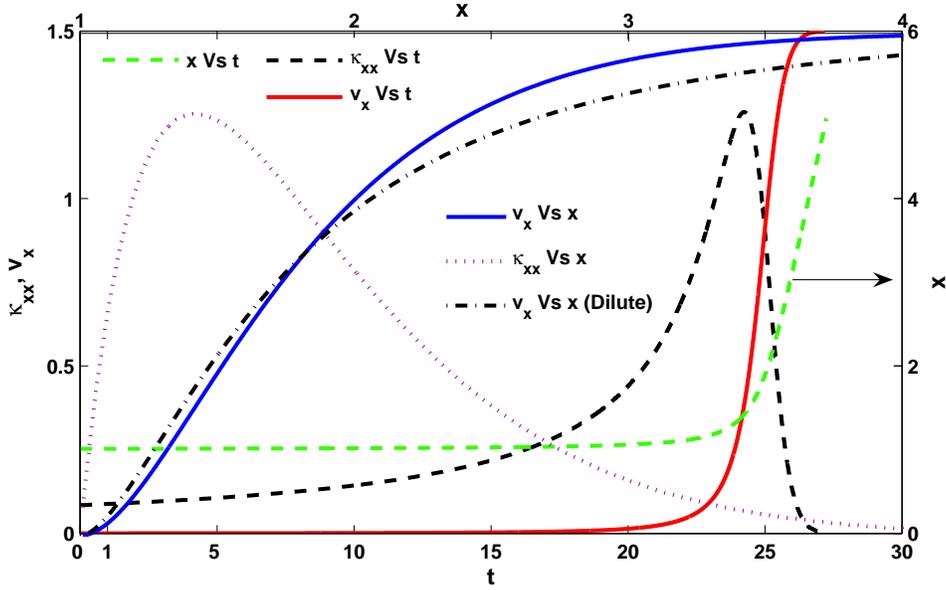}
\end{center}
\caption{(Color online) Velocity $(v_x)$ and velocity gradient
$(\kappa_{xx})$ for an ultra-dilute solution (or a Newtonian fluid) along the center line in the wake of the cylinder,
computed using the FEM formulation with the M5 mesh. The dashed
green line is the position $x$, as a function of time, of a material
particle traveling downstream starting close to the stagnation
point. The dashed black line is the time dependent velocity gradient
$\kappa_{xx}(t)$ used for carrying out BDS of the ultra-dilute models. 
The dot-dashed line is the velocity profile for a dilute Oldroyd-B model at $\Wi = 0.6$.} \label{Vel-Vel-Grad}
\end{figure}

The velocity field $v_x(x)$ and the velocity gradient $\kappa_{xx}
(x)$ along the centerline, computed using the FEM formulation with
the M5 mesh, are displayed in Fig.~\ref{Vel-Vel-Grad}. Note the
quadratic nature of the velocity close to the stagnation point. The
position $x$ of a fluid packet as a function of time $t$, calculated
using eqn~(\ref{tvsx}), is  displayed as  the dashed green line.
Consistent with the boundary conditions for the ODE  for
$\text{M}_{xx} $ above, we use $\delta = 0.01$. Clearly, a
material particle spends a significant fraction of time close to the
stagnation point before rapidly accelerating away from it. The time
dependent velocity gradient $\kappa_{xx}(t)$, necessary for the
integration of the stochastic differential equation~(\ref{sde}),
calculated as described above, is shown as the dashed black line.

The configurational distribution function for the Hookean dumbbell
model is Gaussian both at equilibrium, and in the presence of a
homogenous flow field~\cite{BAH2}. Since a Gaussian distribution is
completely determined by its second moments, the initial
distribution function at $x =1.01$ can be calculated for the known
velocity gradient $\kappa_{xx}$, using the analytical solution for
$\text{M}_{xx}$ in eqn~(\ref{mxxanal}) (and a similar expression for
$\text{M}_{yy}$. Note $\text{M}_{zz}$ = 1). For Hookean dumbbells,
the SDE~(\ref{sde}) is integrated forward in time here, using a
second order predictor-corrector Brownian dynamics simulation (BDS)
algorithm~\cite{Ottingerbook}, with an initial ensemble of connector
vectors distributed according to the Gaussian distribution at $x =
1.01$, subjected to the time dependent velocity gradient
$\kappa_{xx}(t)$ for $t > 0$.

The distribution function is also Gaussian in the case of the FENE-P
model, and as a result, the initial distribution function at $x
=1.01$ can in principle be determined using the second moments
obtained by solving the set of governing equations~(\ref{FENE-P}).
However, we adopt the simpler procedure of using a Gaussian
distributed initial ensemble with equilibrium second moments,
$\text{M}_{xx} =1$, $\text{M}_{yy} =1$, and $\text{M}_{zz} =1$, at
$x=1.01$, since the solution of the SDE downstream of the stagnation
point is found to be insensitive to the choice of the initial
distribution. (For instance, identical results are obtained if the
initial distribution of Hookean dumbbells at $x=1.01$, is used
instead). The BDS algorithm for FENE-P dumbbells is identical to the
one used to integrate the SDE for Hookean dumbbells, with the
additional feature of having to evaluate $\average{{Q^\dagger}^2}$
at every time step.

The exact (numerical) results along the centerline in the cylinder
wake, for an ultra-dilute solution, calculated by solving the ODEs
and the SDE above, are compared with the FEM solution in
section~\ref{RandD} below. Before doing so, however, important
insights can be obtained by considering the nature of the maximum in
the $\text{M}_{xx}$ component in the wake of the cylinder.

\subsection{The maximum in the wake} \label{maxwak}

For an ensemble of dumbbells which start near the stagnation point
(where the velocity gradient $\kappa_{xx}$ is negligibly small), and
then travel downstream to the region of fully developed flow (where
$\kappa_{xx} = 0$), the $\text{M}_{xx}$ component of the
conformation tensor initially has a value close to unity,  and
ultimately returns to a value of unity. Since  $\kappa_{xx} > 0$ at
intermediate values of $x$, it is clear that $\text{M}_{xx}$ must
attain a maximum at some point $x=x^*$ along the centerline in the
wake of the cylinder. Indeed all computations of flow around a
confined cylinder report the occurrence of a maximum, and as
mentioned earlier, failure to attain mesh convergence is usually
observed most noticeably at the maximum.

At the maximum (where $d\text{M}_{xx}/{dx} = 0$), eqn~(\ref{old-B})
implies that,
\begin{equation} \label{maxima}
\text{M}_{xx}|_{x=x^*}=\frac{1}{1-2\, \lambda  \kappa_{xx}|_{x=x^*}}
\end{equation}
Clearly, if $\lambda  \kappa_{xx} = 0.5$ at $x=x^*$, the maximum
value of $\text{M}_{xx}$ will be unbounded. One can see from the
dotted curve in  Fig.~\ref{Vel-Vel-Grad} that for values of $\lambda
= \mathcal{O} (1)$, there are many points along the centerline in
the cylinder wake where $\lambda  \kappa_{xx}$ can be greater than
0.5.  However,  as manifest from eqn~(\ref{maxima}), the real issue
is whether $\lambda  \kappa_{xx} = 0.5$ at $x=x^*$. This question is
examined in the section below, using the different solutions methods
discussed above.
\begin{figure}[tbp]
\begin{center}
\includegraphics[bb= 86 261 508 580, scale=0.75]{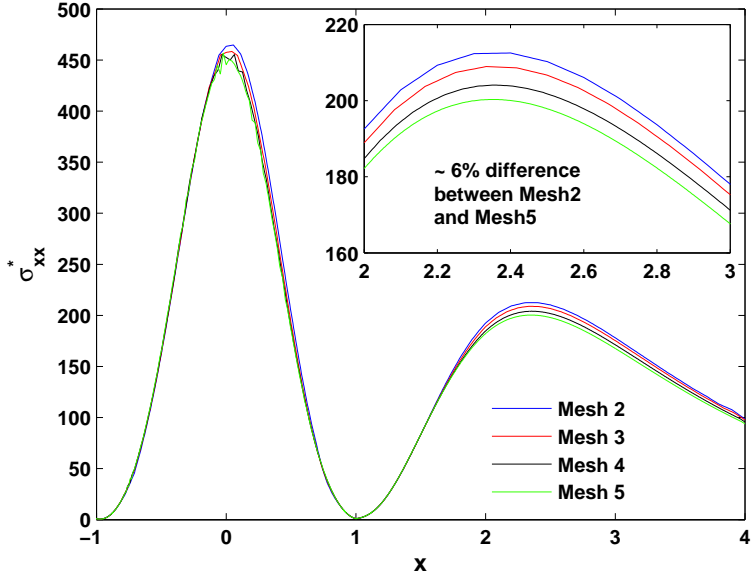}
\end{center}
\caption{(Color online) The dimensionless polymer contribution to
stress, $\sigma^*_{xx}=(\lambda/\eta_{\text{p,0}})\sigma_{xx}$,
along the cylinder wall and along the centerline in the cylinder
wake, at $\Wi=1.0$, for an ultra-dilute solution of an Oldroyd-B
fluid. The computations are carried out using the FEM formulation
discussed in section~\ref{bound}. The inset shows the lack of mesh
convergence in the wake. } \label{Mesh-Divergence}
\end{figure}

\section{Results and Discussion} \label{RandD}

\subsection{Ultra-dilute solutions} \label{ultdilsol}

The failure to attain mesh convergence in the cylinder wake, for an
ultra-dilute polymer solution at $W\!i = 1$, using the FEM
formulation discussed in section~\ref{bound}, is displayed in terms
of the non-dimensional polymer contribution to the stress in
Fig.~\ref{Mesh-Divergence}. The profile for the stress, with two
maxima, one on the cylinder wall, and a second in the wake is
typical for viscoelastic flow around a confined cylinder. As $W\!i$
increases further, the maximum in the wake grows significantly, and
becomes the more dominant of the two maxima. The lack of mesh
convergence, which is already apparent at  $W\!i = 1$ in
Fig.~\ref{Mesh-Divergence}, becomes much more  pronounced. These
predictions are completely in accord with what has been observed
previously for dilute solutions. They have only been reproduced here
to demonstrate the existence of a similar mesh convergence problem
even in the simpler case of an ultra-dilute solution.

\begin{figure}[tbp]
\begin{center}
\begin{tabular}{cc}
\includegraphics[bb= 86 262 508 580, scale=0.5]{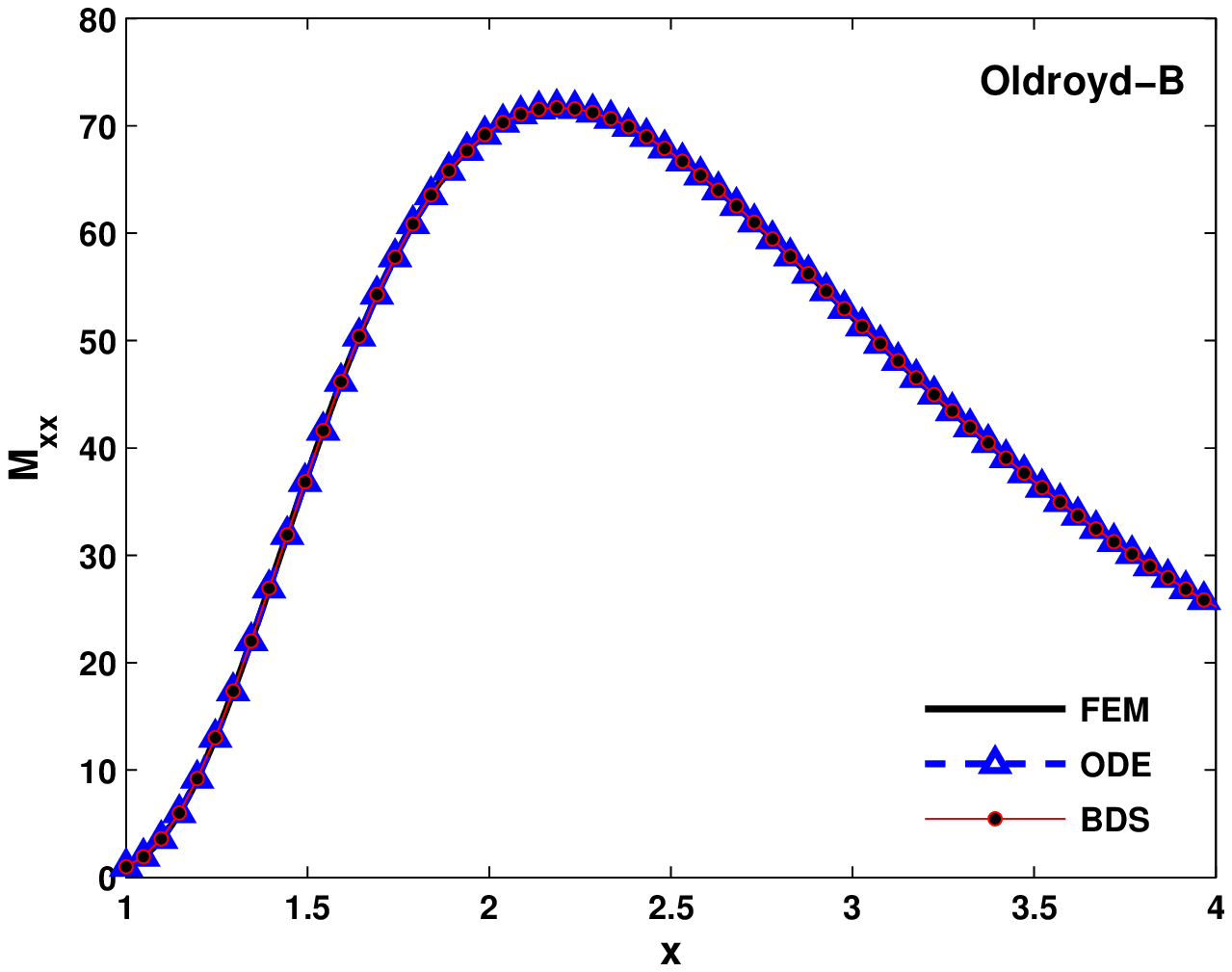} &
\includegraphics[bb= 86 262 508 580, scale=0.5]{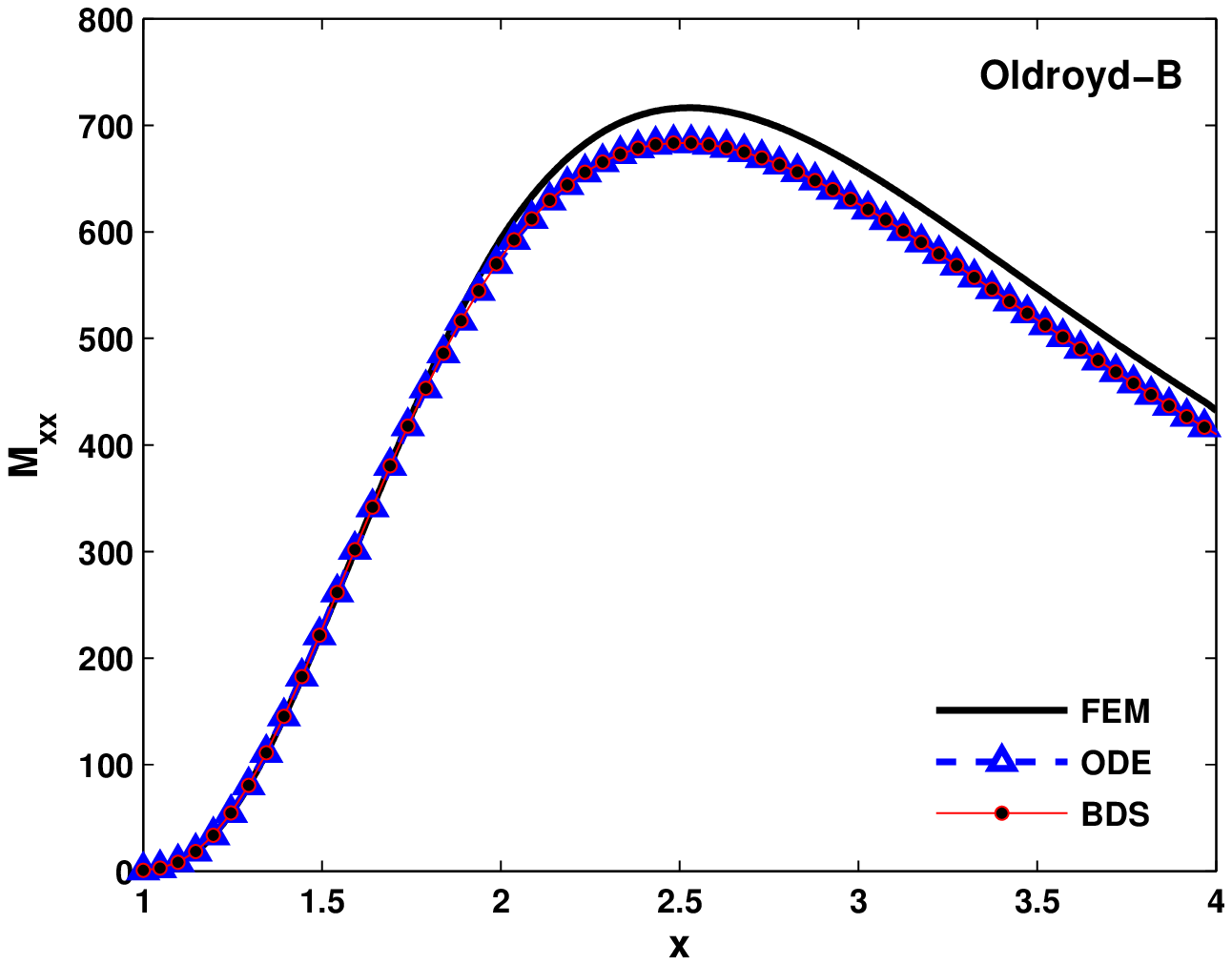}\\
(a) $\Wi=0.8$ & (b) $\Wi=1.3$ \\
\includegraphics[bb= 86 262 508 580, scale=0.5]{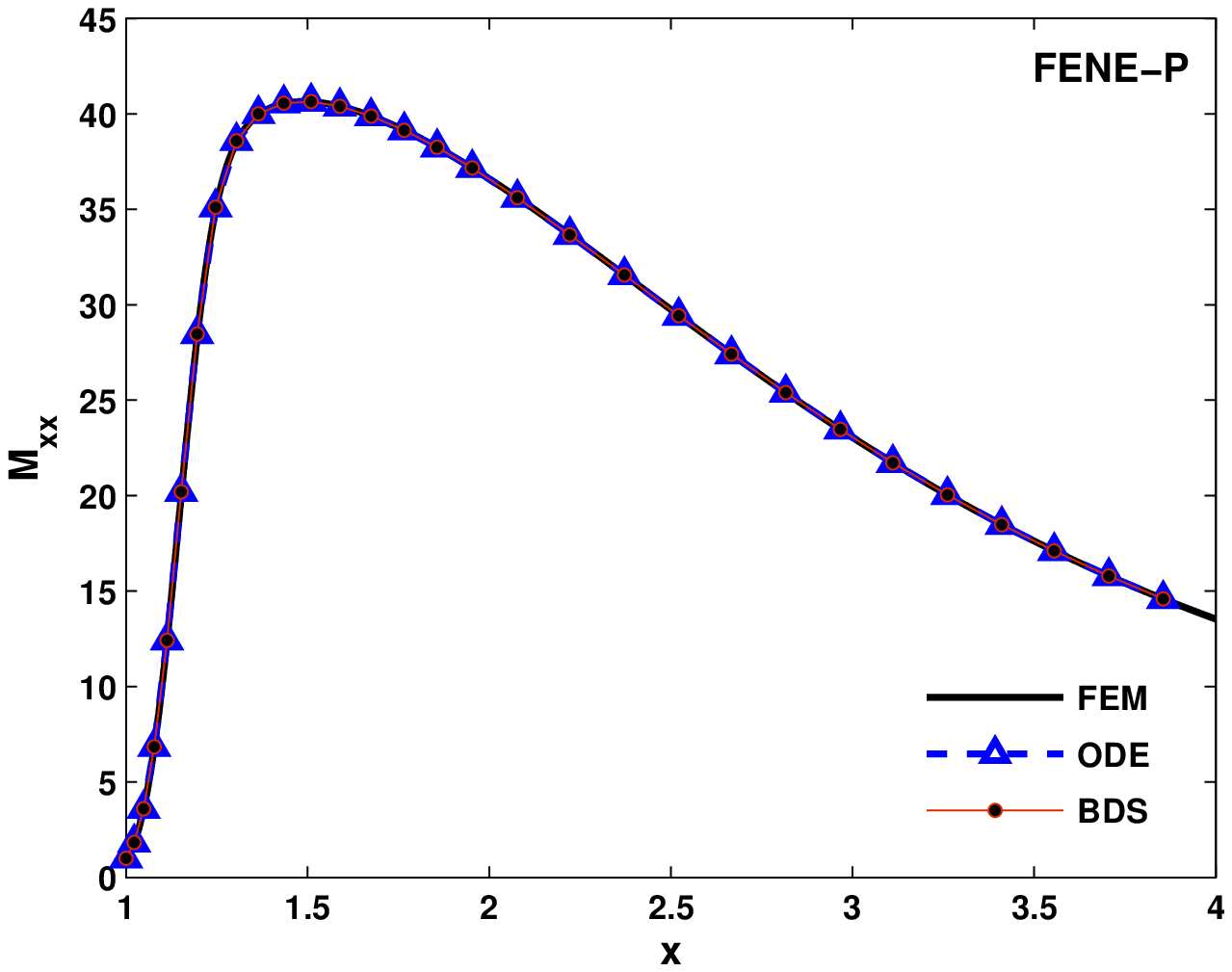} &
\includegraphics[bb= 86 262 508 580, scale=0.5]{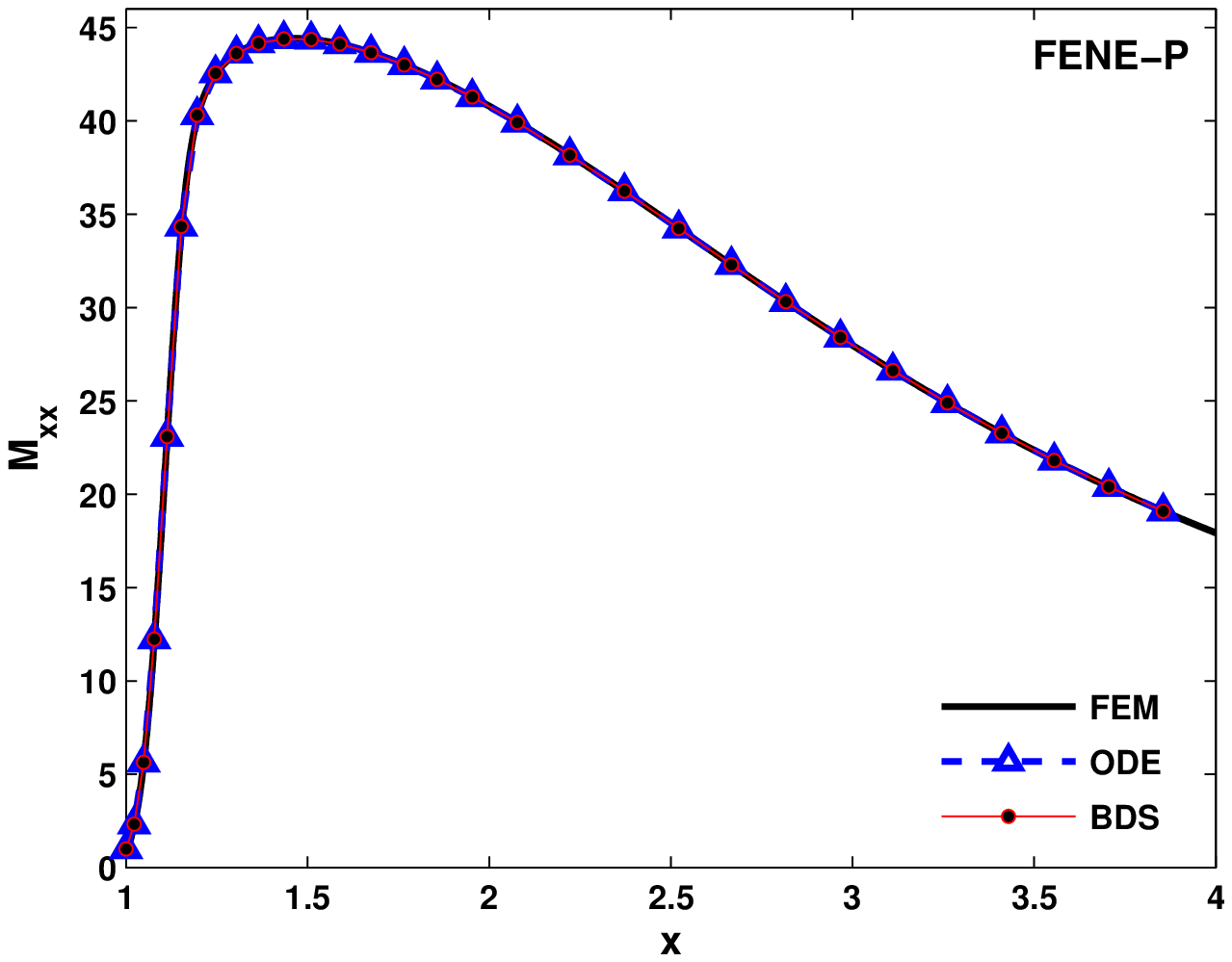}\\
(c) $\Wi=1.5$ & (d) $\Wi=2.0$
\end{tabular}
\end{center}
\caption{(Color online) The $\text{M}_{xx}$ component of the
conformation tensor, computed using FEM, ODE and BDS, for an
ultra-dilute solution of Oldroyd-B and FENE-P ($b=55$) fluids. The
BDS results are obtained by averaging over $10^6$ dumbbell
trajectories. The FEM results are computed using the M5 mesh. For
the Oldroyd-B model, at $W\!i=1.3$, the FEM results at the maximum
are approximately $6\%$ different from the ODE or BDS values.}
\label{fig:MeshConverg-Old-B}
\end{figure}

The advantage of considering ultra-dilute solutions is that exact
(numerical) solutions along the centerline in the cylinder wake can
be obtained as outlined in section~\ref{exacsol}.
Fig.~\ref{fig:MeshConverg-Old-B} compares the predictions of FEM
computations with the exact solutions obtained by solving the ODEs,
eqns~(\ref{old-B}) and (\ref{FENE-P}), and by integrating  (using
BDS) the SDE, eqn~(\ref{sde}), for the Oldroyd-B and FENE-P models.
The excellent agreement of the FEM results with the exact solution
(except for the Oldroyd-B model at $W\!i=1.3$) shows that the FEM
results are accurate at the Weissenberg numbers that have been
displayed.  The departure of FEM predictions, at  $W\!i=1.3$,  from
the exact results for the Oldroyd-B model
(Fig.~\ref{fig:MeshConverg-Old-B} (b)), is clear proof of the
breakdown of FEM computations for $W\!i >1$. Since the exact
solution is known for any value of $W\!i$, the error in the FEM
computation can be estimated. Before we discuss the error, however,
we first consider the more pressing issue of the value of the
non-dimensional strain rate  $\lambda  \kappa_{xx}$ at $x=x^*$, the
location of the maximum in $\text{M}_{xx}$ in the cylinder wake.

\begin{figure}[tbp]
\begin{center}
\begin{tabular}{cc}
\includegraphics[bb= 100 236 490 594, scale=0.6]{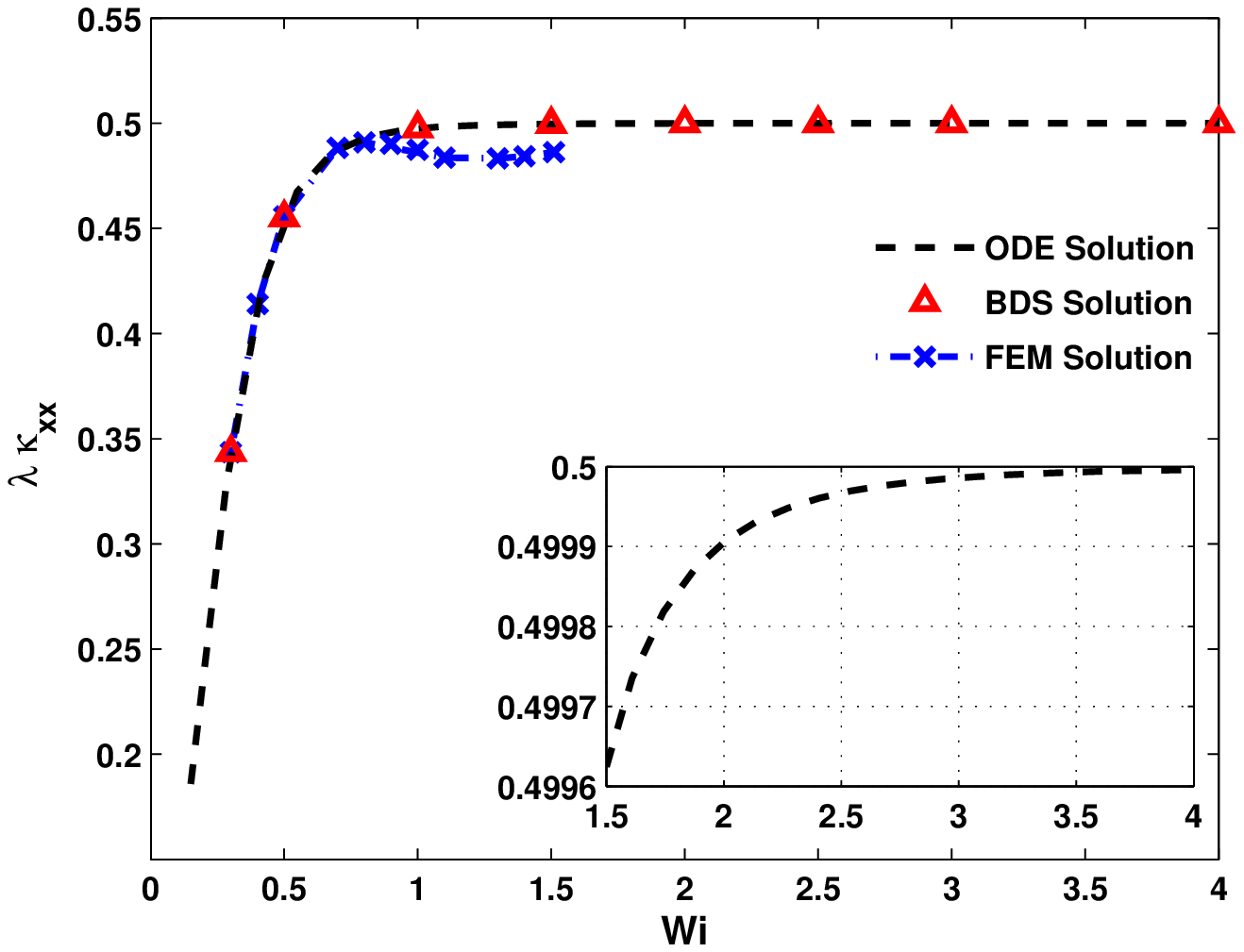}&
\includegraphics[bb= 100 236 490 594, scale=0.6]{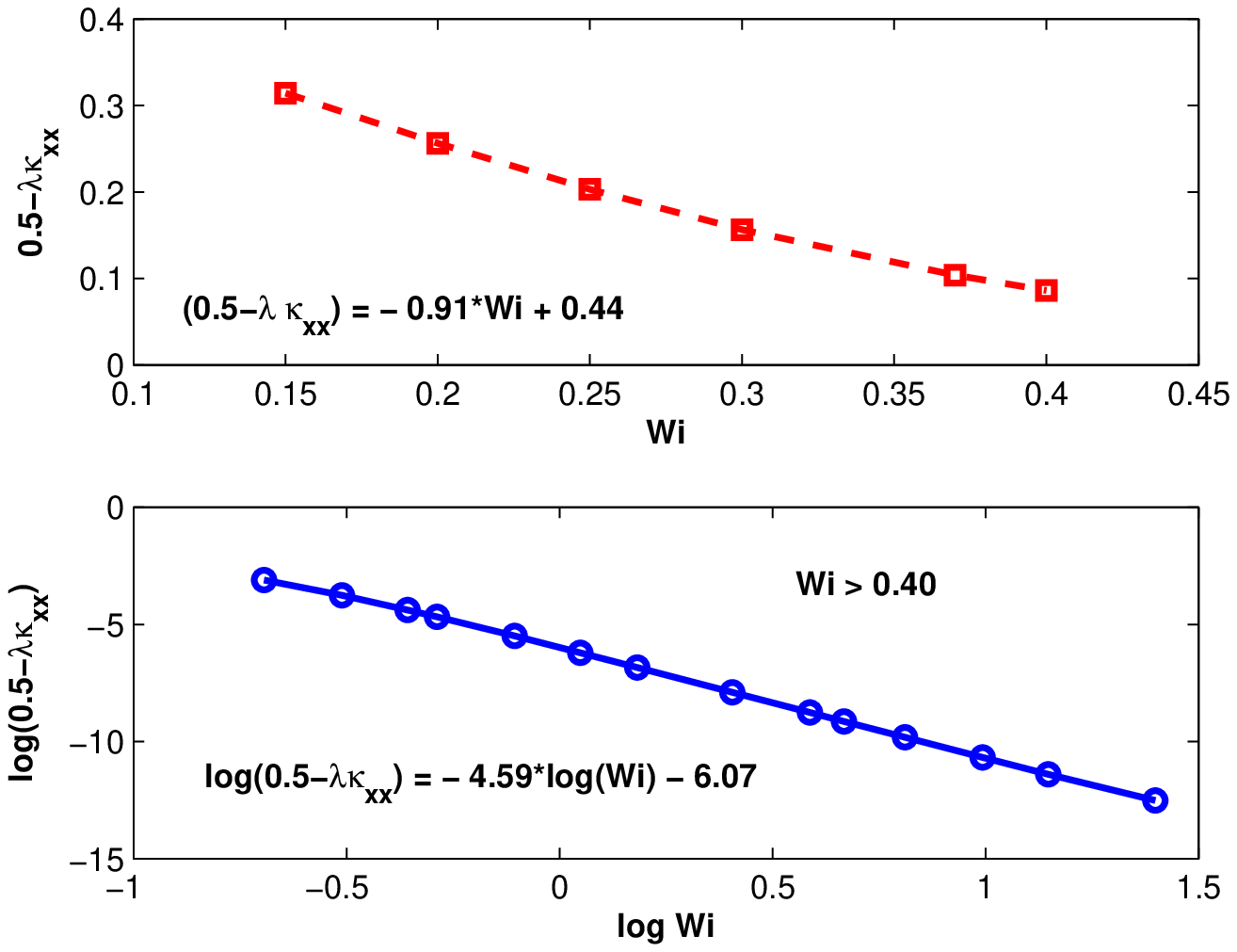}\\
(a)  & (b) \\
\end{tabular}
\end{center}
\caption{(Color online) (a) Dependence of the non-dimensional strain
rate $\lambda\kappa_{xx}$ on $W\!i$, at the location $x=x^*$ of the
maximum in $\text{M}_{xx}$ in the cylinder wake, for an ultra-dilute
Oldroyd-B fluid. Inset shows $\lambda\kappa_{xx}|_{x=x^*}$
approaching $0.5$, computed from the ODE solution, for $\Wi>1.5$.
(b) $\lambda\kappa_{xx}|_{x=x^*}$ approaches the critical value 0.5
as a power-law with increasing $\Wi$. The FEM results are for the M5
mesh and BDS results are obtained by averaging over $10^6$
individual Brownian trajectories of Hookean dumbbells. }
\label{fplambda}
\end{figure}

Fig.~\ref{fplambda}~(a) displays $\lambda\kappa_{xx}|_{x=x^*}$ as a
function of $W\!i$, for an ultra-dilute Oldroyd-B fluid, obtained
with the three different solution techniques. As was discussed
earlier in section~\ref{maxwak}, the maximum in $\text{M}_{xx}$
becomes unbounded as $\lambda\kappa_{xx}|_{x=x^*} \to 0.5$ (see
eqn~(\ref{maxima})). Interestingly, $\lambda\kappa_{xx}|_{x=x^*} $
first approaches $0.5$  at $\Wi \approx 1$, where computational
difficulties with the FEM method are first encountered. While the
ODE solution and BDS can be continued to higher $\Wi$, FEM
computations (indicated by the crosses) are no longer accurate
beyond $\Wi = 1$, breaking down completely by $\Wi = 1.55$. This is
related, as will be discussed in greater detail shortly, to the
inability of the FEM solution to resolve the very large stresses
that arise as $\lambda\kappa_{xx}|_{x=x^*}$ approaches 0.5.
Fig.~\ref{fplambda}~(b) shows that the approach of
$\lambda\kappa_{xx}|_{x=x^*}$ to the critical value is approximately
linear in Weissenberg number for small values of $\Wi$, becoming a
power-law ($\Wi^{-4.59}$), for $\Wi \gtrsim 0.4$. Thus, though
$\lambda\kappa_{xx}|_{x=x^*} \to 0.5$ asymptotically, it never
equals or exceeds the critical value. This implies that
$\text{M}_{xx}$ (and, consequently, $\sigma^*_{xx}$) will increase
without bound as $\Wi$ increases, but will never become singular.

\begin{figure}[tbp]
\begin{center}
\includegraphics[bb= 86 261 508 580, scale=0.8]{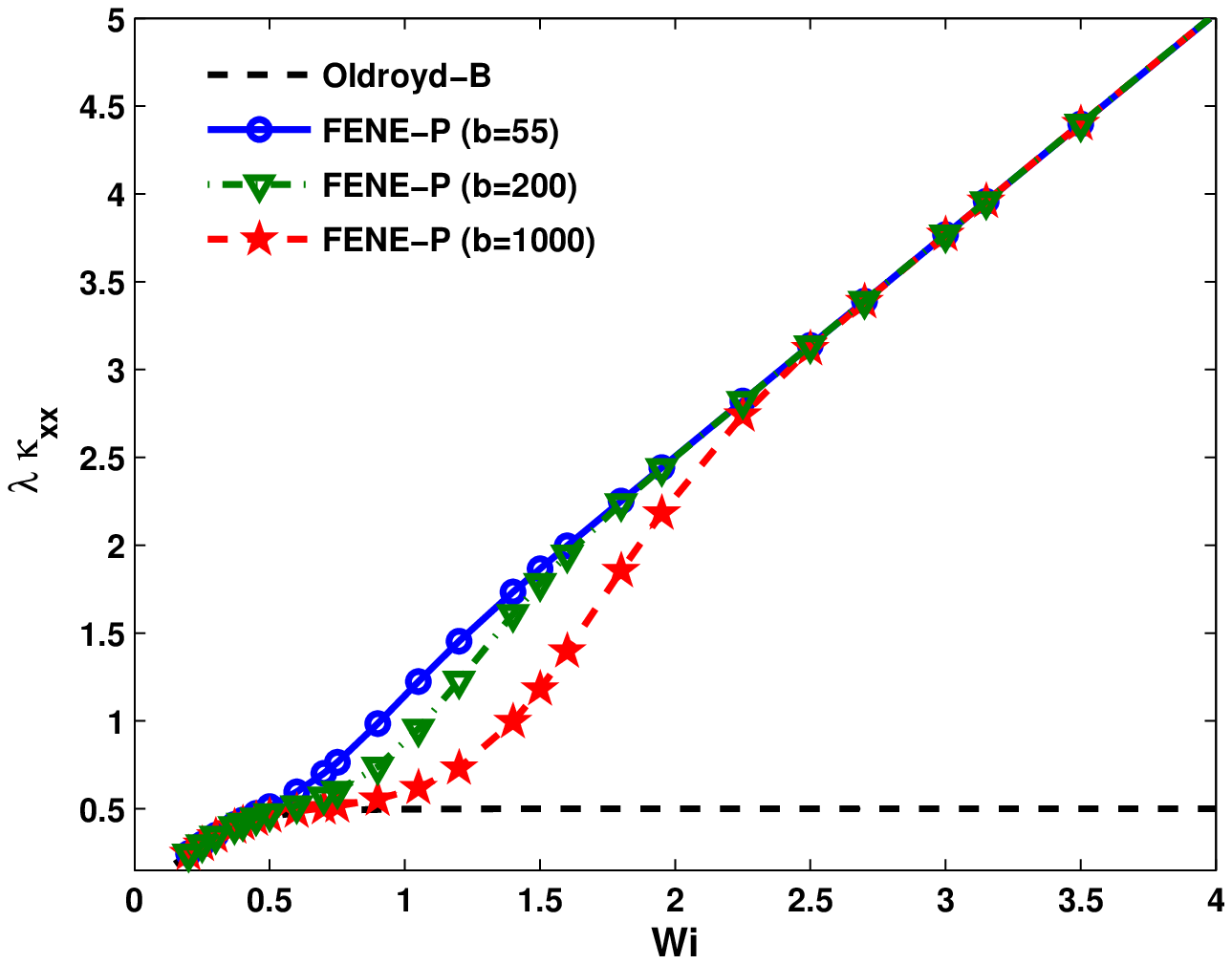}
\end{center}
\caption{(Color online) Dependence of the non-dimensional strain
rate $\lambda\kappa_{xx}$, at the location $x=x^*$ of the maximum in
$\text{M}_{xx}$ in the cylinder wake, on $W\!i$, for a FENE-P
fluid.}\label{fplambdafene}
\end{figure}

In the case of the FENE-P model, $\lambda\kappa_{xx}|_{x=x^*}$
approaches and exceeds  0.5 with increasing $W\!i$, as shown in
Fig.~\ref{fplambdafene}, for a range of values of the
finite-extensibility parameter $b$. The value 0.5 is not significant
for the FENE-P model since, as can be seen from eqn~(\ref{FENE-P}),
$\text{M}_{xx}|_{x=x^*}\rightarrow (b -\tr \dyd{M}/{3})/(\tr
\dyd{M}/{3}-1)$ as $\lambda\kappa_{xx}|_{x=x^*} \to  0.5$. Hence,
$\text{M}_{xx}|_{x=x^*}$ remains finite as long as $b$ is finite.
Before we compare the predictions of $\text{M}_{xx}|_{x=x^*}$ and
$\sigma^*_{xx}|_{x=x^*}$ by the Oldroyd-B and FENE-P models as a
function of $\Wi$, it is interesting to consider the dependence of
the location of the maximum, $x=x^*$, and $\kappa_{xx}|_{x=x^*}$, on
$\Wi$.

\begin{figure}[tbp]
\begin{center}
\begin{tabular}{cc}
\includegraphics[bb= 100 236 490 594, scale=0.6]{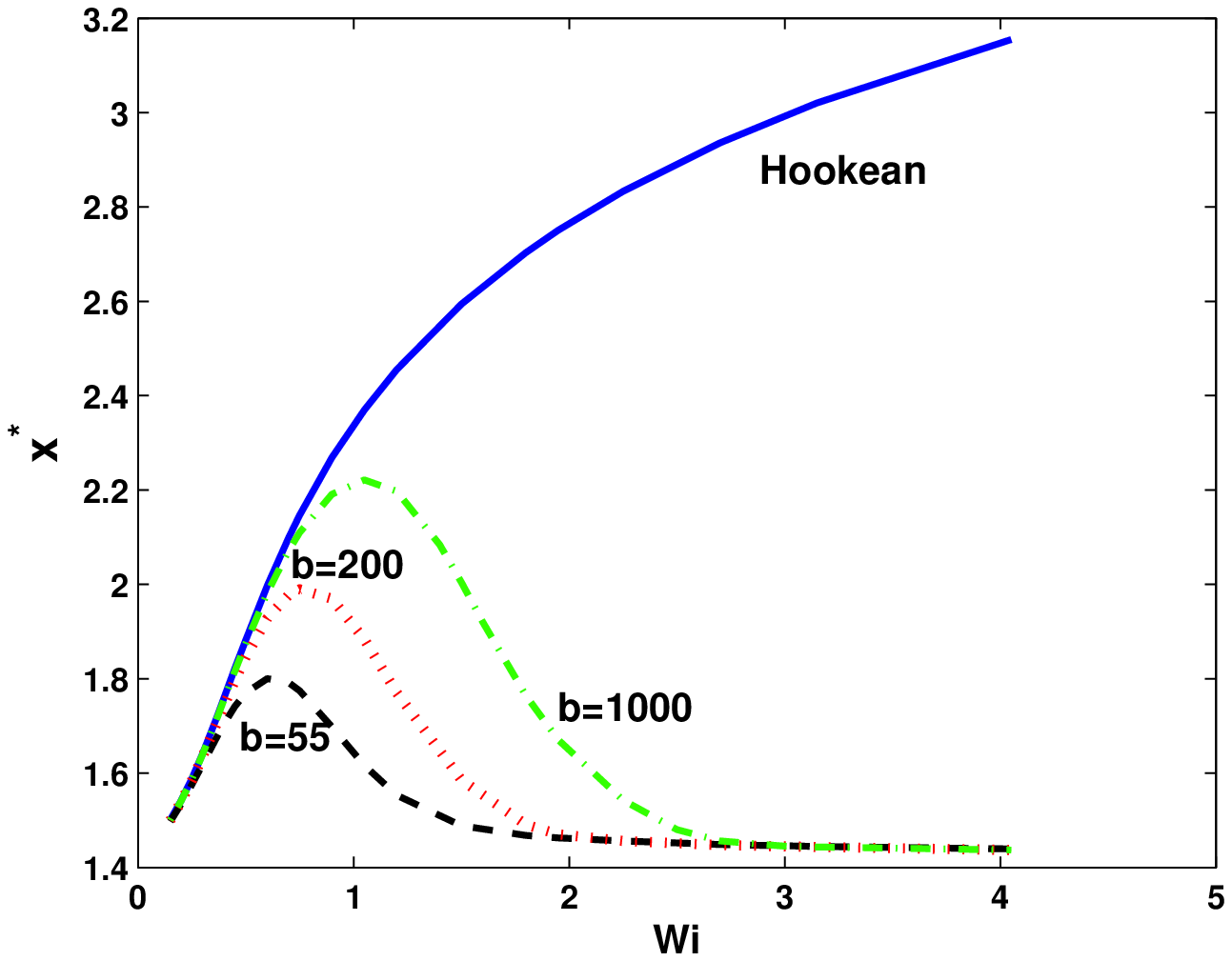} &
\includegraphics[bb= 100 236 490 594, scale=0.6]{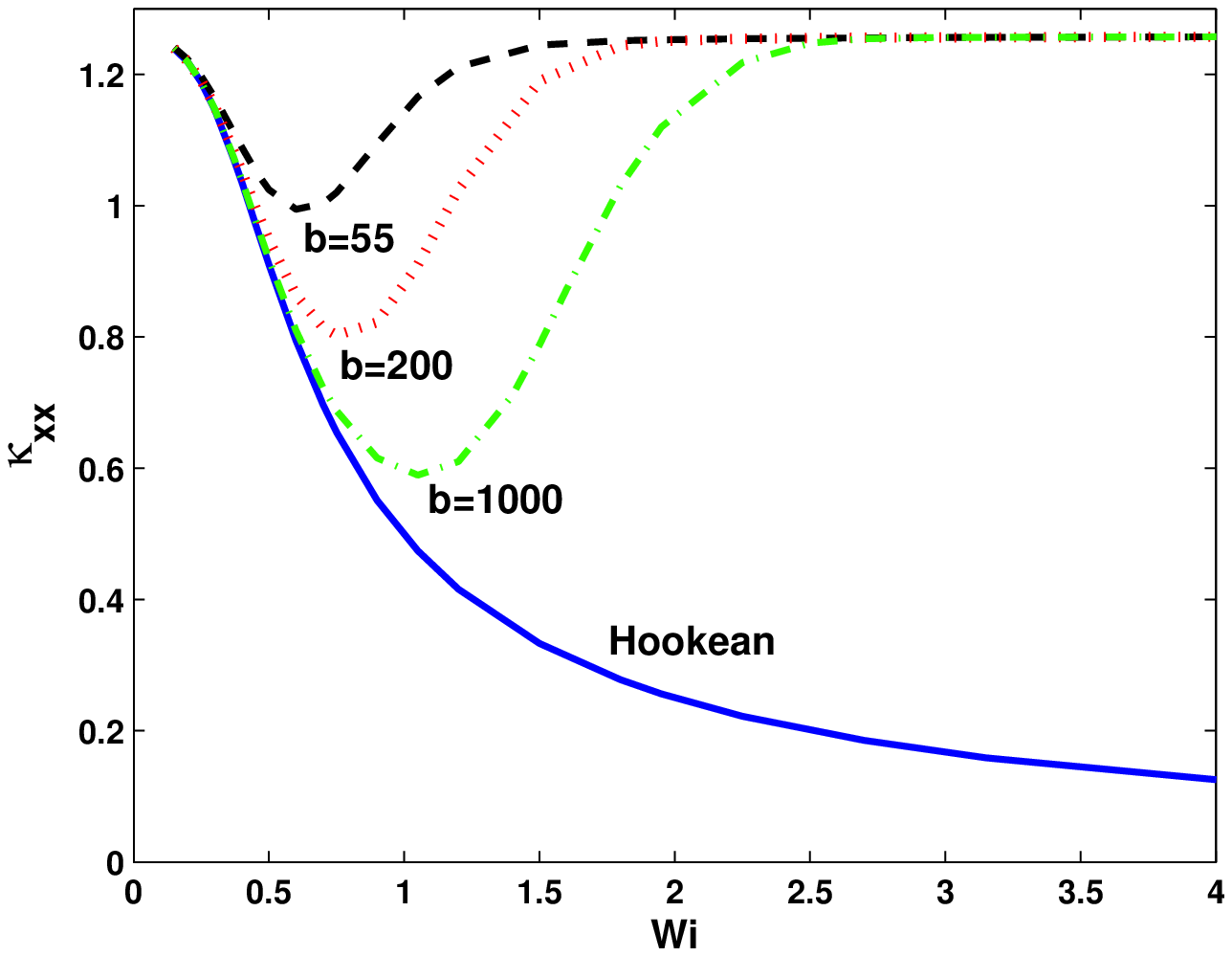}\\
(a) & (b)
\end{tabular}
\end{center}
\caption{(Color online) The dependence on $\Wi$ of (a) the location
$x^*$ of the maximum in $\text{M}_{xx}$, and (b) the strain rate
$\kappa_{xx}|_{x=x^*}$, for ultra-dilute solutions of Oldroyd-B and
FENE-P models. Results displayed are solutions of the respective
ODEs.} \label{X-KXWi}
\end{figure}

Figure~\ref{X-KXWi}~(a) shows that for the Oldroyd-B model, the
location $x^*$ of the maximum in $\text{M}_{xx}$ continuously moves
downstream away from the stagnation point in the cylinder wake, with
increasing $\Wi$. Simultaneously, Fig.~\ref{X-KXWi}~(b) indicates
that the value of  $\kappa_{xx}|_{x=x^*}$ continuously decreases.
This is consistent with the behavior of  $\kappa_{xx}$ as a function
of $x$ displayed in Fig.~\ref{Vel-Vel-Grad}. Since the velocity
field is determined a priori for an ultra-dilute solution, $\Wi$ is
varied in the computations by varying $\lambda$. With increasing
$\Wi$, the product $\lambda\kappa_{xx}|_{x=x^*}$ (with $\lambda$
increasing, and $\kappa_{xx}|_{x=x^*}$ decreasing) tends to 0.5 in
the manner depicted above in Fig.~\ref{fplambda}~(a). The continued
use of the Oldroyd-B fluid in computational rheology, in spite of
its known shortcoming in extensional flow, has sometimes been
justified by the argument that real flows adapt to avoid an infinite
stress. Curiously, the results in Figs.~\ref{fplambda}
and~\ref{X-KXWi} appear to suggest that the flow does regulate the
stress, and avoid a singularity. This does not, however, prevent the
failure of the FEM computations for the Oldroyd-B fluid!

For a FENE-P fluid, the location of the maximum in $\text{M}_{xx}$
and the value of $\kappa_{xx}$ at $x=x^*$, display intriguing
behavior with increasing $\Wi$, as shown in Fig.~\ref{X-KXWi}. For
each value of $b$, beyond some threshold value of $\Wi$, both
quantities attain constant values.  As a consequence, beyond this
threshold value, the product $\lambda\kappa_{xx}|_{x=x^*}$ increases
linearly with $\Wi$, as can be seen clearly in
Fig.~\ref{fplambdafene}, enabling a straightforward mapping between
$\lambda\kappa_{xx}|_{x=x^*}$ and $\Wi$ to be made.

\begin{figure}[tbp]
\begin{center}
\begin{tabular}{cc}
\includegraphics[bb= 100 236 490 594, scale=0.6]{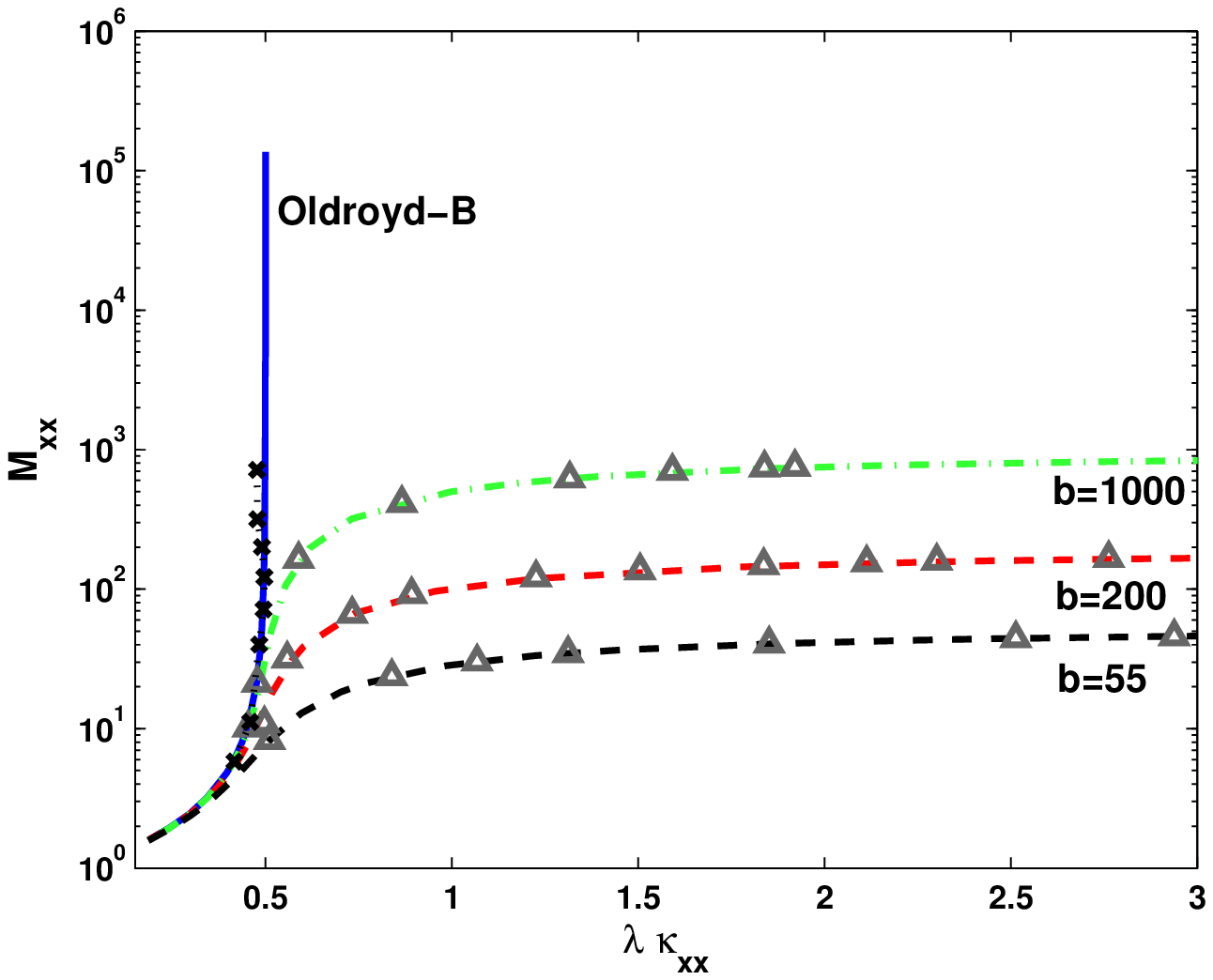} &
\includegraphics[bb= 100 236 490 594, scale=0.6]{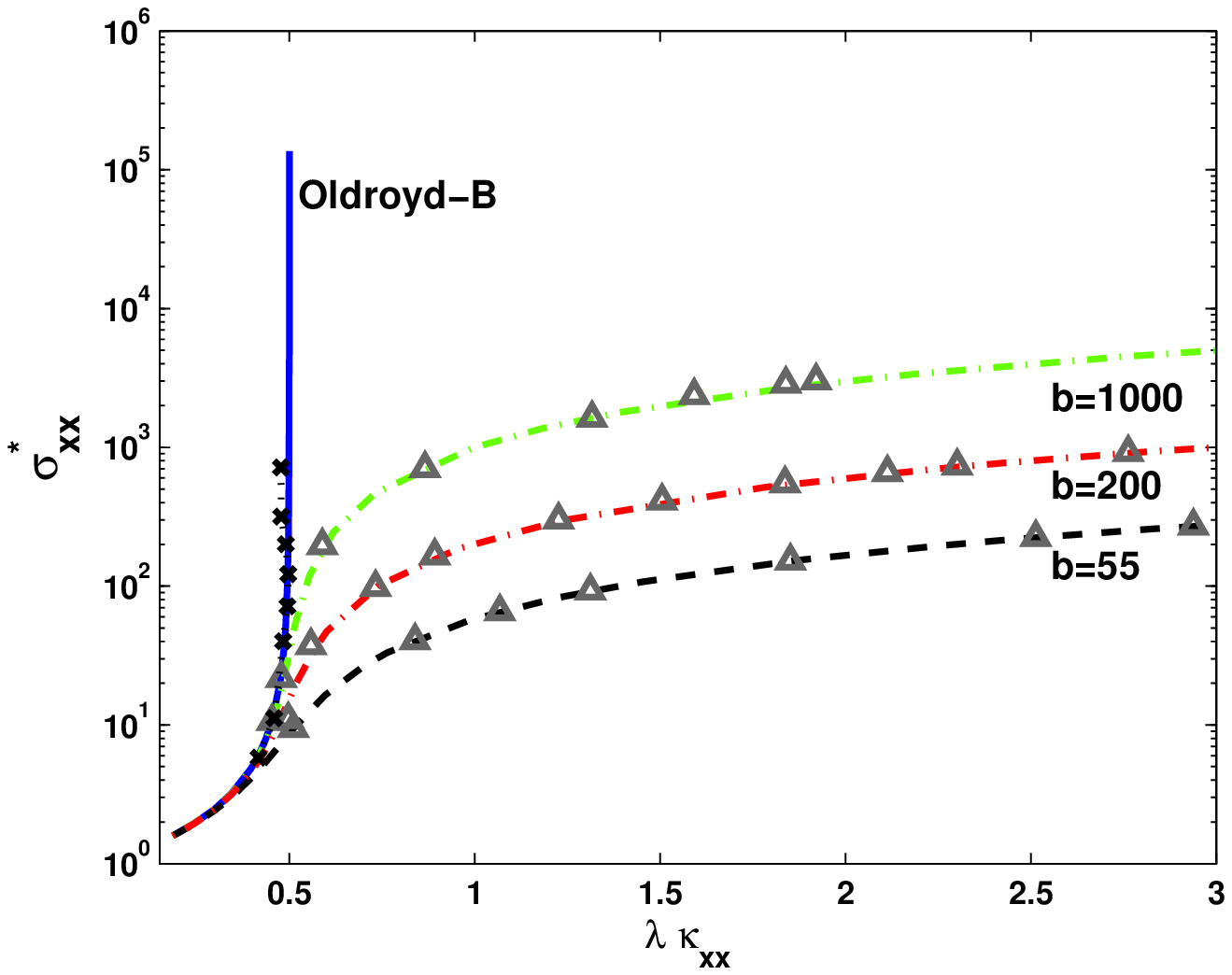}\\
(a) & (b)\\
\end{tabular}
\end{center}
\caption{(Color online) Coil-stretch transition in the cylinder wake
for an ultra-dilute solution. Dependence of the polymer stretch
($\text{M}_{xx}$) and stress ($\sigma^*_{xx}$), at $x=x^*$, on
$\lambda\kappa_{xx}|_{x=x^*}$. The lines are ODE solutions and the
symbols are FEM results on mesh M5.} \label{CoilStretch}
\end{figure}

The steep increase in $\text{M}_{xx}|_{x=x^*}$ and
$\sigma^*_{xx}|_{x=x^*}$ for the Oldroyd-B model, as
$\lambda\kappa_{xx}|_{x=x^*} \to  0.5$, is displayed in
Figs.~\ref{CoilStretch} (a) and (b). For the FENE-P model, there is
a point of inflection at $\lambda\kappa_{xx}|_{x=x^*} =  0.5$, after
which the curves increase much more gradually with increasing
$\lambda\kappa_{xx}|_{x=x^*}$. The shapes of the curves for the
Oldroyd-B and FENE-P models are strikingly reminiscent of the
well-known extensional viscosity versus strain rate curves for these
models, commonly used to display the unphysical behavior of the
Oldroyd-B model~\cite{BAH2,Owensbook}. As is well known, in that
case, the onset of the steep increase in stress is attributed to the
occurrence of a coil-stretch transition, leading to an unbounded
stress in the Oldroyd-B model, but a bounded stress in the FENE-P
model. While the former is because of the infinite extensibility of
the Hookean spring in the Hookean dumbbell model, the latter is
because of the existence of a upper bound to the mean stretchability
of the spring in the FENE-P model. We can conjecture, consequently,
that in the present instance also, polymer molecules undergo a
coil-stretch transition in the wake of the cylinder, at the location
of the stress maximum, giving rise to a stress that increases
without bound as $\Wi$ increases.

\begin{figure}[tbp]
\begin{center}
\begin{tabular}{cc}
\includegraphics[bb= 100 236 490 594, scale=0.6]{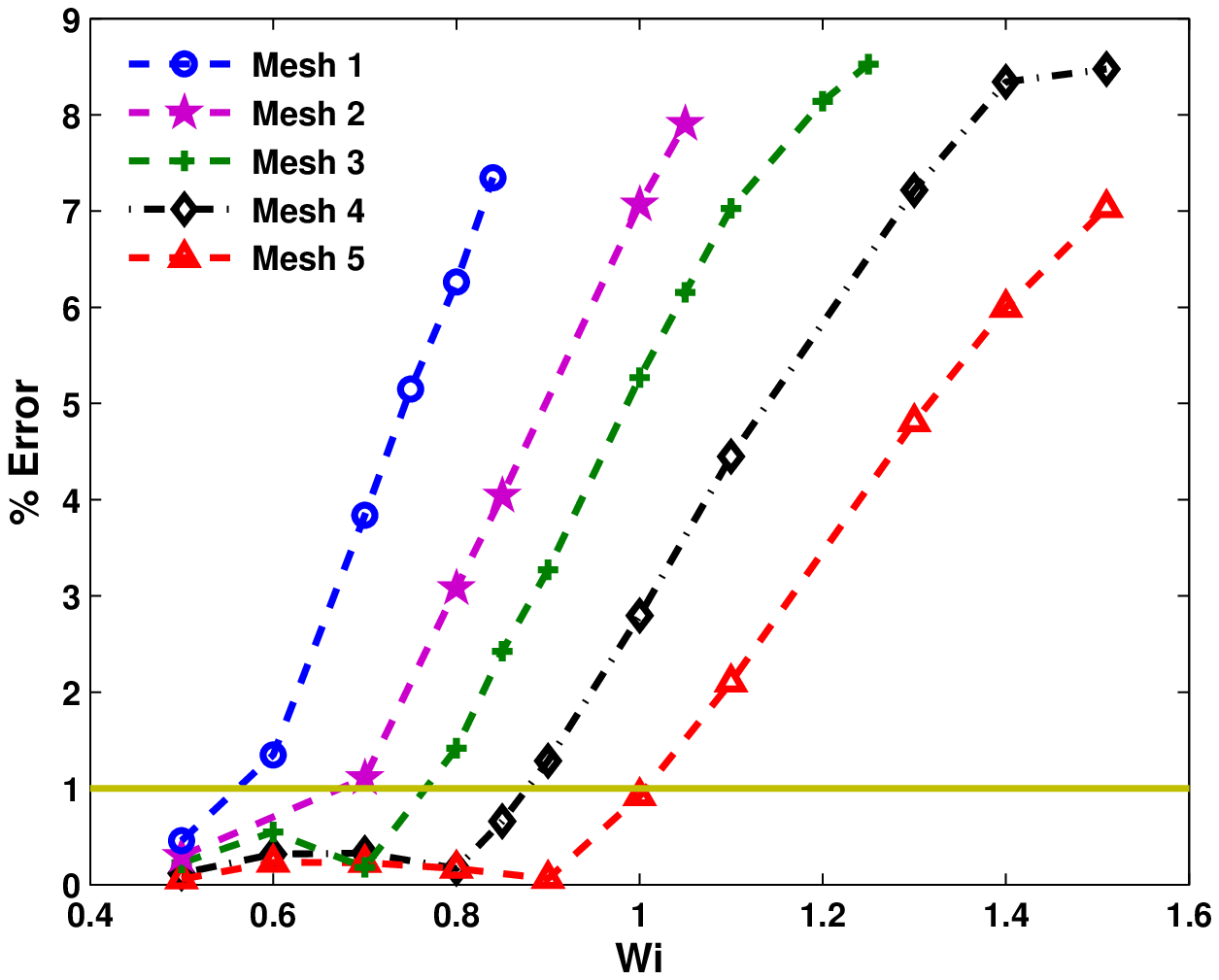} &
\includegraphics[bb= 100 236 490 594, scale=0.6]{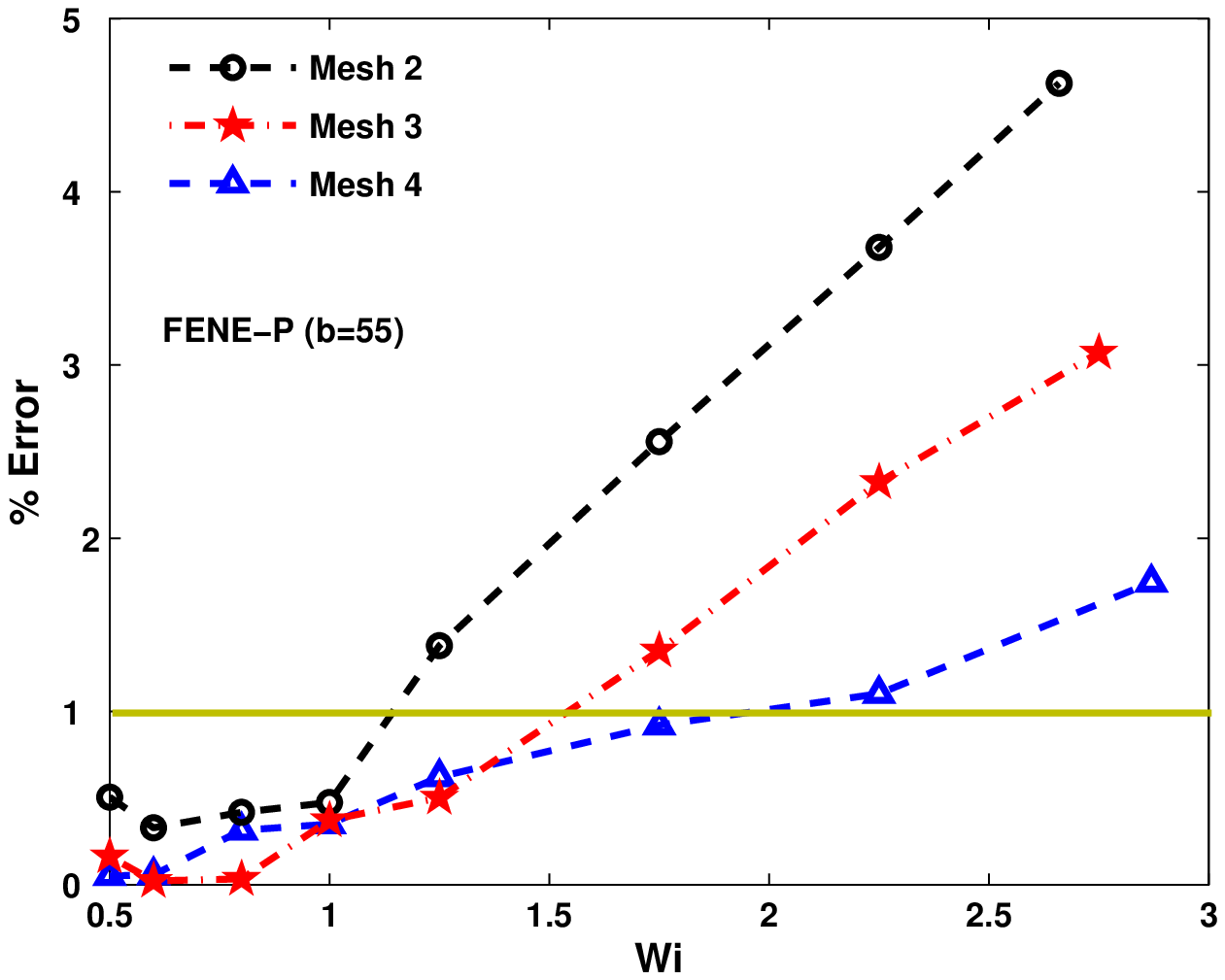}\\
(a) Oldroyd-B & (b) FENE-P \\
\end{tabular}
\end{center}
\caption{(Color online) Percentage error in FEM computations of (a)
$\text{M}_{xx}|_{x=x^*}$ for an ultra-dilute  Oldroyd-B solution,
and (b) $\text{M}_{yy}|_{x=x^*}$ for an ultra-dilute FENE-P fluid,
as a function of $\Wi$. The error is calculated based on the results
of the ODE solutions.} \label{error}
\end{figure}

The breakdown of the FEM computations for the Oldroyd-B model is
clearly related to the steep increase in $\text{M}_{xx}|_{x=x^*}$
and $\sigma^*_{xx}|_{x=x^*}$, as $\lambda\kappa_{xx}|_{x=x^*} \to
0.5$. (Fig.~\ref{CoilStretch}~(b) indicates that the stress maximum
increases by five orders of magnitude as $\Wi$ increases from 0.1 to
4). Since the exact solution is known at $x=x^*$, we can calculate
the error at any value of $\Wi$ using,
\begin{equation}
\text{error}=\frac{(M^{\text{ODE}}_{xx}|_{x=x^*}-M^{\text{FEM}}_{xx}|_{x=x^*})}{M^{\text{ODE}}_{xx}|_{x=x^*}}\times
100 \label{erroreqn}
\end{equation}
Error estimates obtained in this manner are displayed in
Fig.~\ref{error}~(a) for the Oldroyd-B model. The error remains
small  $(<1\%)$ at low $\Wi$, but increases sharply to approximately
$8\%$ as $\Wi \gtrsim \mathcal{O} (1)$. The value of $\Wi^*$, the
Weissenberg number up to which the error is less than $1\%$, depends
on the degree of mesh refinement, and a clear improvement in $\Wi^*$
can be observed with increased mesh refinement. However, to obtain
mesh converged results for $\Wi>0.7$ (with error $<1\%$), an
approximately 100 fold increase in mesh density and hence,
approximately $\sim 100$ fold increase in computational time is
required.

In the case of the FENE-P model, the error in
$\text{M}_{xx}|_{x=x^*}$ is small even at relatively large values of
$\Wi$ since the chains are close to their fully extended length, and
it is difficult to see a clear pattern in the change in error with
mesh refinement, unlike in the case of the Oldroyd-B model above.
However, the error in the $\text{M}_{yy}|_{x=x^*}$ component reveals
a more systematic behavior, as displayed in Fig.~\ref{error}~(b),
with a decrease in error with increasing mesh refinement.

\begin{figure}[tbp]
\begin{center}
\includegraphics[bb= 100 236 490 594, scale=0.7]{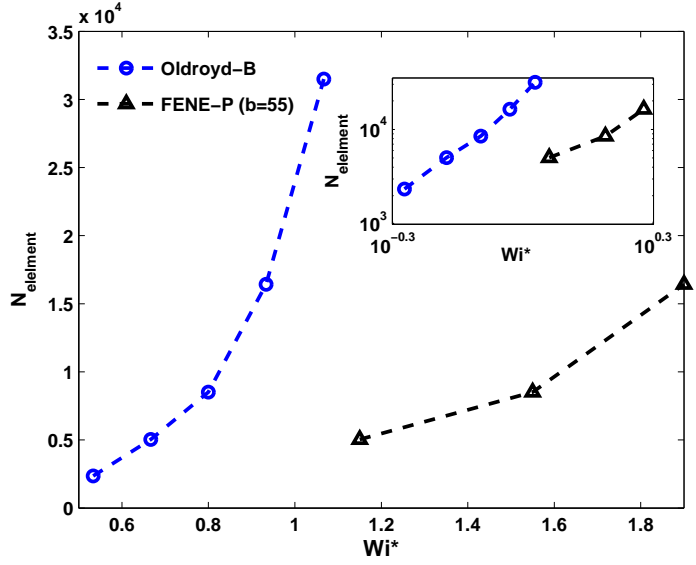} 
\end{center}
\caption{(Color online) The number of elements as a function of the Weissenberg
number $\Wi^*$, up to which the error remains less than $1\%$, for the
Oldroyd-B and FENE-P models. The inset displays the same data as a log-log plot.} \label{nelement}
\end{figure}

The infeasibility of carrying out FEM computations for the confined
flow around a cylinder of an ultra-dilute Oldroyd-B fluid, for $\Wi
> 1$, is revealed in Fig.~\ref{nelement}, where the
exponential increase in the number of elements required to attain an
error less than $1\%$, with increasing $\Wi$, can be clearly
observed. For the FENE-P model, the rate of increase in the number
of elements required for an error less than $1\%$ is significantly
lower than that for the Oldroyd-B fluid. The Mesh 4 curve in
Fig.~\ref{error}~(b) even seems to suggest that there might be a
degree of mesh refinement beyond which, for the FENE-P model, one
can compute at any $\Wi$ with an error less than $1\%$. However,
this trend is not easily discernible in Fig.~\ref{nelement}, and
addtional mesh refinement may be required before a firm conclusion can
be drawn. Further, a change of variable to the matrix logarithm of
the conformation tensor, may lead to mesh converged results at
significantly higher vales of $\Wi$.

\begin{figure}[tbp]
\begin{center}
\includegraphics[bb= 86 261 508 580, scale=0.7]{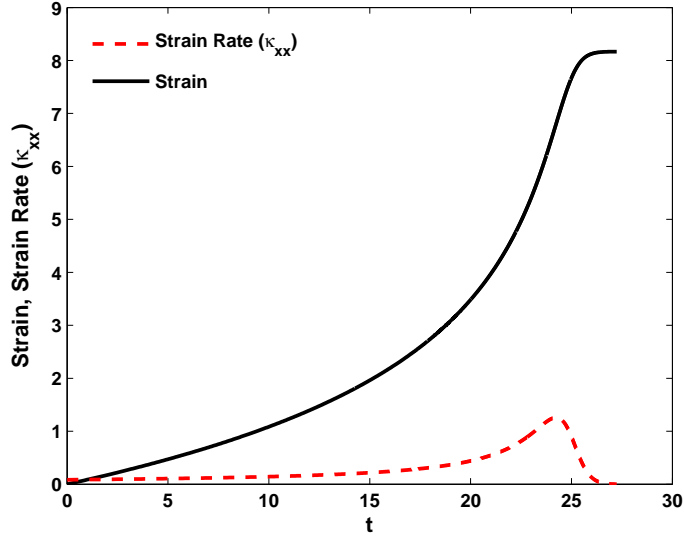}
\end{center}
\caption{(Color online) Hencky strain accumulated by a fluid element
as it travels along the centerline, from the
neighborhood of the stagnation point in the cylinder
wake.}\label{strain}
\end{figure}

The reason polymer molecules undergo a coil-stretch transition as
they travel down the centerline in the cylinder wake is because of
the extended period of time they spend in the neighborhood of the
stagnation point, which leads to a significant accumulation of
strain. The Hencky strain $\epsilon$ at any instant $t$, calculated
from the expression $\epsilon = \int_{0}^t dt^\prime \, \kappa_{xx}
(t^\prime) $, is displayed in Fig.~\ref{strain}. A very large Hencky
strain of roughly 8 units is built up by the time the molecules
approach $x^*$. For a material element of unit length at $t=0$, this
corresponds to a ratio of final to initial length of roughly 3000.
The behavior of individual molecules as they are subjected to this
degree of straining can be obtained, for an ultra-dilute solution,
from the Brownian dynamics simulations carried our here since one
can calculate the trajectories of dumbbells as they are convected by
the flow field down the centerline, subjected to the local strain
rate.

In this context, it is instructive to calculate the size of
individual dumbbells relative to a macroscopic feature, such as the
length of an element in the finite element mesh. In the
non-dimensionalization scheme used here, however, since the
macroscopic length scale is the cylinder radius $a$, and the
microscopic length scale for Hookean dumbbells is
$\sqrt{\average{Q^2}_\text{eq}/3}$, direct comparison is difficult
unless one has estimates of these length scales. Here, we use the
experimental data of~\citet{McKinley1993}, who investigated the flow
around a confined cylinder (with radius $3.188 \times 10^{-3} \,
\text{m}$) of a 1.2 million molecular weight Polyisobutelene
solution, to obtain a typical estimate of these length scales. For
these molecules, the equilibrium size can be shown to be
$\sqrt{\average{Q^2}_\text{eq}} = 0.0497\, \mu\text{m}$.
Defining a dimensionless length in the flow direction by,
\begin{equation}
Q^*_x=\frac{Q^\dagger_{x} \, \sqrt{\average{Q^2}_\text{eq}/3}}{a \,
L_m|_{x=x^*}}
\end{equation}
where, $L_m|_{x=x^*}$ is the non-dimensional length of the element
at $x=x^*$, the relative length of individual molecules in the flow
direction can be calculated from the BDS trajectories as a function
of strain.

\begin{figure}[tbp]
\begin{center}
\includegraphics[bb= 100 236 490 594, scale=0.8]{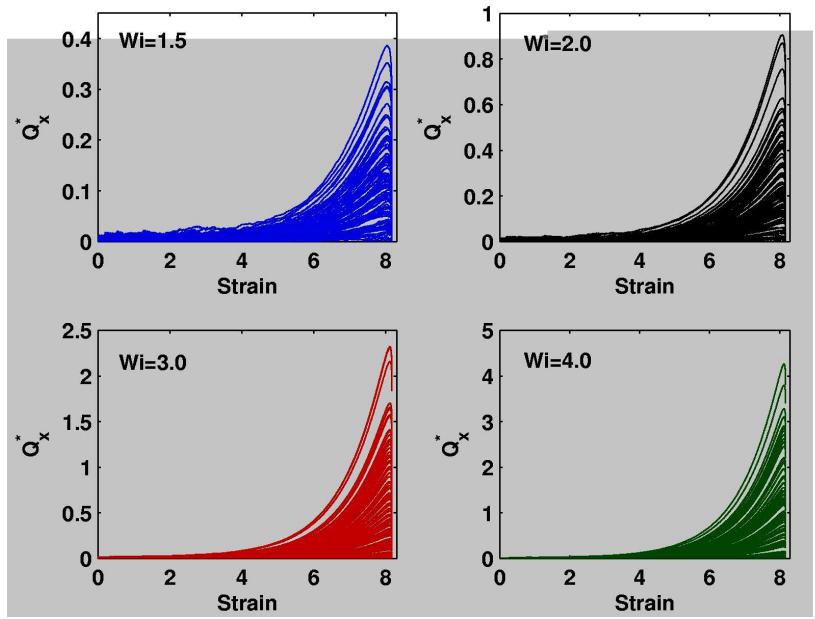}\\
\end{center}
\caption{(Color online) The length of individual polymer molecules
relative to the size of an element of the M5 mesh, at $x=x^*$, at
various values of $\Wi$, for an ultra-dilute Oldroyd-B fluid. The
experimental data of~\citet{McKinley1993} is used to obtain an
estimate of the equilibrium size of the molecules and the cylinder
radius.}\label{Qxelement}
\end{figure}

Figures~\ref{Qxelement} display $Q^*_x$ as a function of $\epsilon$,
for an ensemble of 100 dumbbell trajectories, at various values of
$\Wi$. Nearly all the dumbbells appear to remain close to their
initial state of extension until approximately 5 strain units,
beyond which several of the dumbbells undergo rapid extension, which
is more pronounced as $\Wi$ increases. The rapid extension of the
dumbbell spring represents the physical unraveling of a polymer
molecule from a coiled to a stretched state, and the results in
Figs.~\ref{Qxelement} are inline with the notion that a coil-stretch
transition occurs as the molecules experience the maximum strain. As
expected, the molecules relax back to their equilibrium
configurations once the strain rate downstream of the maximum
becomes zero.

The use of the local element size to achieve non-dimensionalization
reveals strikingly that the magnitude of some of the molecules is
large enough to span several elements. Kinetic theory models, such
as the Hookean dumbbell model, are typically built on the assumption
of homogenous fields, with negligible variation on the length scale
of individual molecules. The data in Figs.~\ref{Qxelement} suggests
that the extensive use of the unphysical Oldroyd-B model in complex
flow simulations is questionable, and highlights the need to derive
more refined models that are valid in non-homogeneous fields.

\begin{figure}[tbp]
\begin{center}
\includegraphics[bb= 100 236 490 594, scale=0.8]{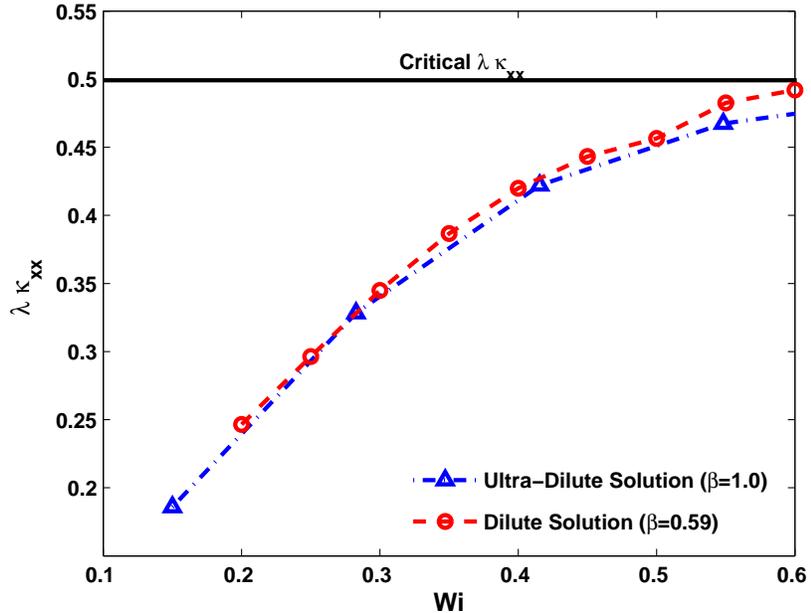}
\end{center}
\caption{(Color online) Dependence of $\lambda\kappa_{xx}|_{x=x^*}$
on $\Wi$ for dilute and ultra-dilute solutions. The dilute solution
curve first approaches the critical value at a value of $\Wi$ at
which difficulties with FEM computations usually arise.}
\label{Dilutelambdakappa}
\end{figure}

\subsection{Dilute solutions} \label{dilsolsec}

All the results reported so far have been for ultra-dilute models,
where the existence of exact solutions has enabled us to obtain a
variety of insights into the origin of difficulties encountered with FEM
computations. There is already an extensive literature on the
numerical computation of the flow around a confined cylinder of
various dilute solution models, and there are no new numerical
techniques introduced in this paper for us to be able to report an
improvement in the maximum attainable mesh converged Weissenberg
number. Our interest here, instead, is to examine if any of the insight
that has been gained for ultra-dilute solutions can be used to
understand the observed behavior of dilute solutions.

The coupling of the velocity and the conformation tensor (and
stress) fields, makes it impossible to obtain exact solutions for
these variables. As a result, it is not possible to obtain error
estimates as in the case of ultra-dilute solutions, or to calculate
the trajectories of individual dumbbell molecules convected along
the centerline by the flow. Nevertheless, we can exploit the key
insight of the previous section because the value of the maximum in
the $\text{M}_{xx}$ component in the cylinder wake, for the
Oldroyd-B model, is still given by eqn.~(\ref{maxima}), even though
we do not have a pre-determined velocity field $v_x(x)$.

\begin{figure}[tbp]
\begin{center}
\includegraphics[bb= 100 236 490 594, scale=0.8]{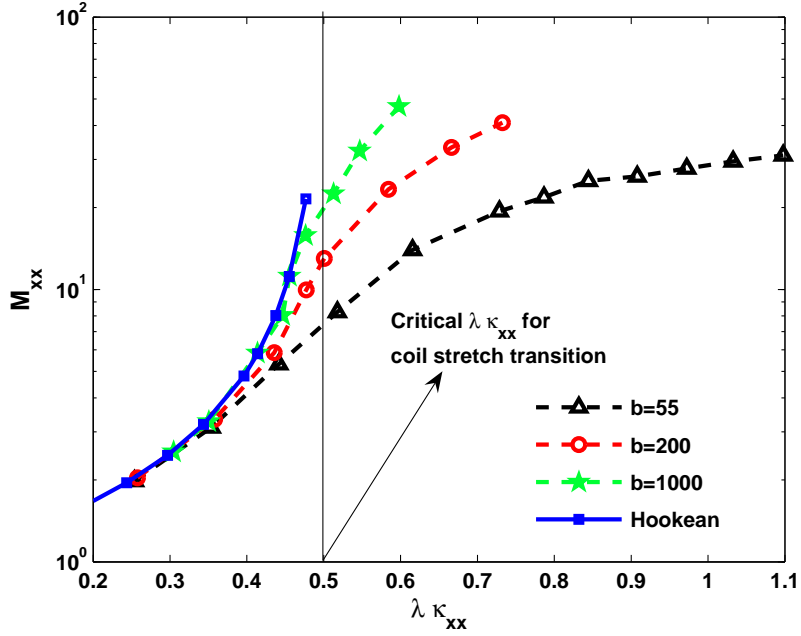}
\end{center}
\caption{(Color online) Coil-stretch transition in the cylinder wake
for a dilute solution. Variation of the polymer stretch
$\text{M}_{xx}$ with $\lambda\kappa_{xx}|_{x=x^*}$ for Oldroyd-B and
FENE-P models. The M4 mesh is used for the Oldroyd-B model and the
M3 mesh is used for the FENE-P model in FEM computations. The
viscosity ratio $\beta=0.59$.} \label{Dilute-CoilStretch}
\end{figure}

In the case of an ultra-dilute solution, it was relatively
straightforward to find the location $x^*$ of the stress maximum in
the cylinder wake for each value of $\Wi$, and to calculate
$\lambda\kappa_{xx}|_{x=x^*}$ from the known velocity field. For a
dilute solution, the velocity field changes with each change in
$\Wi$. All the same, it is still possible to find
$\lambda\kappa_{xx}|_{x=x^*}$, at various values of $\Wi$, by
integrating the full system of equations for the velocity and
conformation tensor fields using the FEM formulation. A typical
velocity profile for a dilute solution at $\Wi=0.6$ is shown in
Fig.~\ref{Vel-Vel-Grad} as the dot-dashed line. (At this value of
$\Wi$, it can be seen that the velocity profile is not significantly
different from that for an ultra-dilute solution.)

Figure~\ref{Dilutelambdakappa} indicates that
$\lambda\kappa_{xx}|_{x=x^*}$ for a dilute solution approaches the
critical value of 0.5 in a manner similar to that of an ultra-dilute
solution, albeit at a slightly more rapid rate with increasing
$\Wi$. The curve is not as smooth as the ultra-dilute case because
of the growing error in FEM computations as $\Wi$ reaches the upper
limit of computable values. Indeed, the typical limit of $\Wi=0.6$
where most computations in the literature first encounter problems,
appears to be the value at which $\lambda\kappa_{xx}|_{x=x^*}$ first
comes close to 0.5.

As may be anticipated from eqn.~(\ref{maxima}), $\text{M}_{xx}$ and
$\sigma^*_{xx}|_{x=x^*}$ will increase steeply for the Oldroyd-B
model as $\lambda\kappa_{xx}|_{x=x^*} \to 0.5$. This can be seen
very clearly from  Fig.~\ref{Dilute-CoilStretch}, where
$\text{M}_{xx}$ appears to become unbounded in this limit. For the
FENE-P fluid on the other hand, the curves for $\text{M}_{xx}$
exhibit a point of inflection at $\lambda\kappa_{xx}|_{x=x^*} =
0.5$, before leveling off to the fully stretched value corresponding
to the respective value of $b$.

The similarity of the shapes of the curves in
Fig.~\ref{Dilute-CoilStretch} to the curves in
Fig.~\ref{CoilStretch}, and to the well-known extensional viscosity
versus strain rate curves for the Oldroyd-B and FENE-P models
suggests that even for a dilute solution, there occurs a
coil-stretch transition at the location of the stress maximum in the
cylinder wake, and this coil-stretch transition is the source of
problems encountered with FEM computations for the flow of an
Oldroyd-B fluid around a confined cylinder. It would be of great
interest to examine if a similar coil-stretch transition is the
source of computational difficulties encountered in the numerical
simulation of other benchmark complex flows of Oldroyd-B fluids.

\section{Conclusions} \label{conc}

The flow around a cylinder confined between parallel plates, of
ultra-dilute and dilute polymer solutions, modeled by the Oldroyd-B
and FENE-P constitutive equations, has been considered with a view
to understand the origin of computational difficulties encountered
in numerical simulations.

FEM computations of ultra-dilute Oldroyd-B solutions are shown to
breakdown at $\Wi = \mathcal{O}(1)$, as has been observed previously
for dilute solutions (see Fig.~\ref{Mesh-Divergence}). Two different
numerical means of obtaining an exact solution along the centerline
in the cylinder wake, for both the Oldroyd-B and the FENE-P models,
have been developed to enable a careful examination of the causes of
the breakdown. The exact solution techniques, namely solving a
system of ODEs and carrying out Brownian dynamics simulations, are
useful to evaluate the value of $\Wi$ up to which the FEM
computations are accurate (see Figs.~\ref{fig:MeshConverg-Old-B}),
and to estimate the error in the FEM results (see
Figs.~\ref{error}).

An analysis of the structure of the Oldroyd-B equation shows that
the maximum in the $\text{M}_{xx}$ component of the conformation
tensor along the centerline in the cylinder wake, becomes unbounded
if the non-dimensional strain rate at the location of the stress
maximum, $\lambda\kappa_{xx}|_{x=x^*}$, approaches a critical value
of $0.5$ (see eqn~(\ref{maxima})). Numerical solution of the
governing equations for an ultra-dilute solution reveals that
$\lambda\kappa_{xx}|_{x=x^*} \to 0.5$ as a power-law in $\Wi$, for
$\Wi \gtrsim \mathcal{O}(1)$ (see Fig.~\ref{fplambda}~(a) and~(b)).
On the other hand, analysis of the FENE-P model reveals that the
maximum in $\text{M}_{xx}$ remains bounded for all values of the
local non-dimensional strain rate. In contrast to the Oldroyd-B
model, numerical results for $\lambda\kappa_{xx}|_{x=x^*}$ show that
it does not have an asymptotic value, but instead increases linearly
with $\Wi$ beyond a threshold value of the Weissenberg number (see
Fig.~\ref{fplambdafene}).

As $\lambda\kappa_{xx}|_{x=x^*} \to 0.5$, both the maximum in
$\text{M}_{xx}$ and the maximum in the stress ($\sigma^*_{xx}$)
increase without bound for the ultra-dilute Oldroyd-B model. In
comparison, for an ultra-dilute FENE-P model, these variables
increase relatively rapidly as $\lambda\kappa_{xx}|_{x=x^*}$
approaches 0.5, but level off and remain bounded for higher values
of $\lambda\kappa_{xx}|_{x=x^*}$ (see Figs.~\ref{CoilStretch}). The
shape of the curves are strongly suggestive of the occurrence of a
coil-stretch transition in the cylinder wake.

The steep increase in $\text{M}_{xx}$ and $\sigma^*_{xx}$ in the
vicinity of $\Wi = \mathcal{O}(1)$ necessitates the use of
increasingly refined meshes for increasing values of $\Wi$. The
number of elements required to maintain the error in $\text{M}_{xx}$
less than 1\% is shown to increase exponentially with increasing
$\Wi$ for the Oldroyd-B model, making it practically infeasible to
obtain solutions at $\Wi > 1$. In the case of the FENE-P model, the
current simulation data is inadequate to draw firm conclusions (see
Figs.~\ref{nelement}).

A material element of an ultra-dilute solution is shown to
accumulate nearly 8 units of Hencky strain as it travels downstream 
from the stagnation point in the wake of the cylinder, due to the
extended time it spends in the vicinity of the stagnation point (see
Fig.~\ref{strain}). This strain leads to dumbbells undergoing a
large extension in the flow direction, with their magnitude large
enough to span several elements in the local finite element mesh see
Fig.~\ref{Qxelement}).

The analysis of the nature of the maximum in $\text{M}_{xx}$ in the
cylinder wake, which suggests that the maximum becomes unbounded if
$\lambda\kappa_{xx}|_{x=x^*}$ approaches the critical value of
$0.5$, is valid for both ultra-dilute and dilute Oldroyd-B fluids.
FEM computations of the fully coupled governing equations for a
dilute Oldroyd-B fluid have been carried out to show that, just as
in the case of an ultra-dilute solution,
$\lambda\kappa_{xx}|_{x=x^*}$ tends to 0.5 with increasing $\Wi$
(see Fig.~\ref{Dilutelambdakappa}). The approach, however, is more
rapid than the ultra-dilute case, with $\lambda\kappa_{xx}|_{x=x^*}$
becoming nearly equal to 0.5 at $\Wi \approx 0.6$, the value of the
Weissenberg number where computational difficulties have been
reported in the literature to be first typically encountered.

The approach of $\lambda\kappa_{xx}|_{x=x^*}$ to the critical value
is shown to be accompanied by an unbounded increase in the maximum
values of $\text{M}_{xx}$ and $\sigma^*_{xx}$ in the cylinder wake
for a dilute Oldroyd-B fluid. FEM computations of the coupled
governing equations for a FENE-P fluid on the other hand show that
these variables increase relatively rapidly close to
$\lambda\kappa_{xx}|_{x=x^*} = 0.5$, but level off and remain
bounded at higher values (see Figs.~\ref{Dilute-CoilStretch}). The
similarity of the curves with observations for ultra-dilute
solutions, is strong evidence for a coil-stretch
transition also occurring in dilute solutions, in the wake of the
cylinder at the location of the stress maximum.

Several issues that must be addressed in the future
can be tackled fruitfully with the framework developed here. For instance,
the nature and structure of stress boundary layers in the vicinity
of the cylinder can be examined for ultra-dilute solutions along
 lines similar to the analysis here. Further, the existence of a
coil-stretch transition suggests that a model with conformation
dependent drag might reveal the existence of coil-stretch hysteresis
in the cylinder wake.

\vskip10pt \noindent \textbf{Acknowledgements} This work has been
supported by a grant from the Australian Research Council under the
Discovery-Projects program, and the National Science Foundation. The
authors would like to thank the APAC, VPAC (Australia) and the RTC
(Rice University, Houston) for the allocation of computing time on
their supercomputing facilities. JRP acknowledges helpful discussions with 
Professors Eric Shaqfeh and Gareth McKinley.

\bibliographystyle{JORnat}

\end{document}